\documentclass{aa}

\usepackage{times,epsfig} 
\usepackage{mathrsfs}  
\usepackage{natbib}
\usepackage{amssymb}
\bibpunct{(}{)}{;}{a}{}{,}
\usepackage{caption}
\usepackage[varg]{txfonts}
\usepackage[section]{placeins}
\usepackage[english]{babel}
\usepackage{url}
\usepackage{hyperref}
\usepackage{arydshln}
\usepackage{stfloats}
\usepackage{color}
\usepackage[normalem]{ulem}

\def\phs{\phantom{$-$}}

\def\kms{{\rm km\,s$^{-1}$}}

\authorrunning{Polshaw et al.}
\titlerunning{LSQ13fn}

\begin{document}

\title{LSQ13fn: A type II-Plateau supernova with a possibly low metallicity progenitor that breaks the standardised candle relation}

\author{
J. Polshaw\inst{1}{\thanks{E-mail: jpolshaw01@qub.ac.uk}}
\and R. Kotak\inst{1}
\and L. Dessart\inst{2}
\and M. Fraser\inst{3}
\and A. Gal-Yam\inst{4}
\and C. Inserra\inst{1}
\and S. A. Sim\inst{1}
\and S. J. Smartt\inst{1}
\and J. Sollerman\inst{5}
\and C. Baltay\inst{6}
\and D. Rabinowitz\inst{6}
\and S. Benetti\inst{7}
\and M. T. Botticella\inst{8}
\and H. Campbell\inst{3}
\and T.-W. Chen\inst{9}
\and L. Galbany\inst{10,}\inst{11}
\and R. McKinnon\inst{6}
\and M. Nicholl\inst{1}
\and K. W. Smith\inst{1}
\and M. Sullivan\inst{12}
\and K. Tak\' ats\inst{10,13}
\and S. Valenti\inst{14}
\and D. R. Young\inst{1}
}

\institute{
Astrophysics Research Centre, School of Mathematics and Physics, Queen’s University Belfast, Belfast BT7 1NN, UK
\and Laboratoire Lagrange, Universit\'e C\^{o}te d'Azur, Observatoire de la C\^{o}te d'Azur, CNRS, Boulevard de l'Observatoire, CS 34229, 06304 Nice Cedex 4, France
\and Institute of Astronomy, University of Cambridge, Madingley Road, Cambridge, CB3 0HA, UK
\and Department of Particle Physics and Astrophysics, Weizmann Institute of Science, Rehovot 76100, Israel
\and Department of Astronomy and the Oskar Klein Centre, Stockholm University, AlbaNova, SE-106 91 Stockholm, Sweden
\and Department of Physics, Yale University, New Haven, CT 06250-8121, USA
\and INAF-Osservatorio Astronomico di Padova, Vicolo dell’Osservatorio 5, I-35122 Padova, Italy
\and INAF-Osservatorio Astronomico di Capodimonte, Salita Moiariello 16, Napoli, 80131 Italy
\and Max-Planck-Institut f{\"u}r Extraterrestrische Physik, Giessenbachstra\ss e 1, 85748, Garching, Germany
\and Millennium Institute of Astrophysics, Vicu\~{n}a Mackenna 4860, 7820436 Macul, Santiago, Chile
\and Departamento de Astronom\'ia, Universidad de Chile, Camino El Observatorio 1515, Las Condes, Santiago, Chile
\and School of Physics and Astronomy, University of Southampton, Southampton SO17 1BJ, UK
\and Departamento de Ciencias F\'isicas, Universidad Andr\'es Bello, Avda. Rep\'ublica 252, 32349 Santiago, Chile
\and University of California Davis, 1 Shields Avenue, Davis, CA 95616
06520
}

\date{Received ... / Accepted ...}

\abstract
{
We present optical imaging and spectroscopy of supernova (SN) LSQ13fn, a type II supernova with several hitherto-unseen properties.
Although it initially showed strong symmetric spectral emission features attributable to \ion{He}{ii}, \ion{N}{iii}, and \ion{C}{iii}, reminiscent of some interacting SNe, it transitioned into an object that would fall more naturally under a type II-Plateau (IIP) classification.
However, its spectral evolution revealed several unusual properties: metal lines appeared later than expected, were weak, and some species were conspicuous by their absence. 
Furthermore, the line velocities were found to be lower than expected given the plateau brightness, breaking the SN~IIP standardised candle method for distance estimates.
We found that, in combination with a short phase of early-time ejecta-circumstellar material interaction, metal-poor ejecta, and a large progenitor radius could reasonably account for the observed behaviour. 
Comparisons with synthetic model spectra of SNe~IIP of a given progenitor mass would imply a progenitor star metallicity as low as 0.1\,Z$_{\odot}$.
LSQ13fn highlights the diversity of SNe~II and the many competing physical effects that come into play towards the final stages of massive star evolution immediately preceding core-collapse.
}

\keywords{supernovae: general -- supernovae: individual: LSQ13fn}

\maketitle

\section{Introduction}\label{sec:intro}

The current generation of transient surveys have led to the discovery of thousands of supernova (SN) candidates per year, with several hundred of these being spectroscopically classified \citep{Gal-Yam_2013}. 
This has allowed studies of sizeable samples of objects, enabling general trends in observed properties to be identified. 
Nevertheless, studies of individual SNe continue to retain a principal role in the literature, especially when they are distinguished by some combination of proximity and peculiarity. 
Indeed, many of these have served to both confirm and challenge currently accepted paradigms \citep[e.g.][for SN~1987A, SN~2011fe, and SN~2009ip, to name but a few]{Arnett_1989b,Nugent_2011,Fraser_2015}.

The current SN spectroscopic classification scheme rightly seeks to identify only the most salient features in an optical spectrum, resulting in a generic breakdown of objects that do (type II) and those that do not (type I) show features attributable to hydrogen \citep{Minkowski_1941,Oke_1974,filippenko_1997}. 
Further divisions are based on the presence or absence of helium in combination with the strength of features such as \ion{Si}{ii}. 
A further spectroscopic distinction is made if symmetric narrow lines ($\lesssim$ few 100s\,\kms; `n' subtype) are present \citep{Schlegel_1990}. 
These narrow lines arise from dense circumstellar material in the vicinity of the progenitor prior to explosion.
Two additional classifications of SNe~II, based on light curve behaviour alone, are also possible -- these are the "II-Plateau (P)" and "II-Linear (L)" varieties that describe the evolution of the lightcurve in the optical bands, at epochs past peak brightness \citep{Barbon_1979}. 
The photometric and spectroscopic classifications are not mutually exclusive; if anything, they can be restrictive when describing objects that morph into other subtypes, or display other peculiarities.
While this system has worked relatively well, the larger volumes surveyed these days continue to yield an ever-increasing number of peculiar events that transcend the traditional classification boundaries.

\begin{figure}[t!]
\centering
\includegraphics[width=1\linewidth,angle=0]{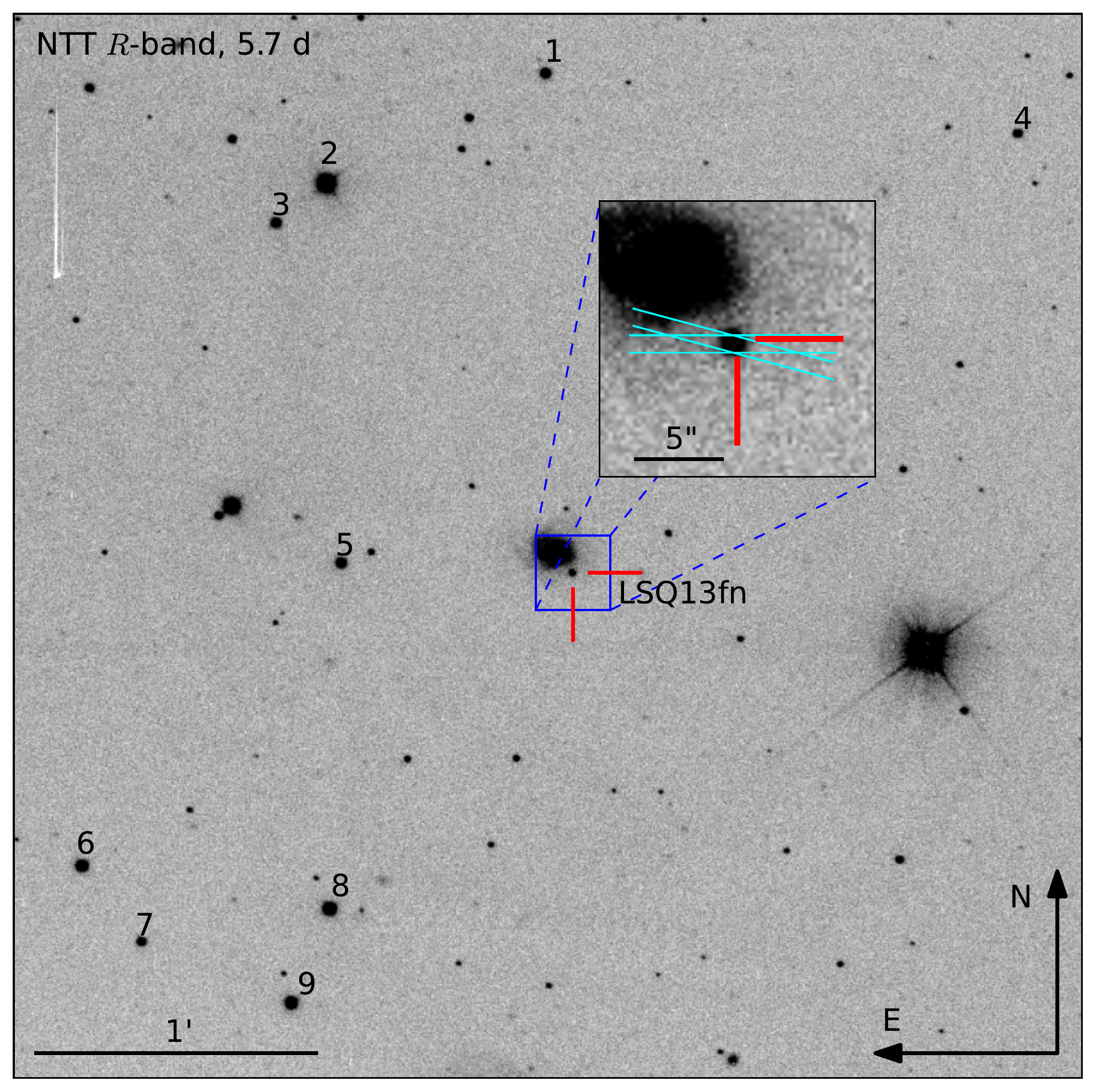}
\caption{
NTT $R$ band image of LSQ13fn, taken at 5.7\,d after maximum light. 
The position of the SN is marked by red dashes, at the coordinates $\alpha$=11:51:17.29, $\delta$=$-$29:36:41.1 (J2000), $4\farcs26$\,W and $4\farcs68$\,S of the nucleus of the host galaxy. The sequence stars used for photometric calibration are indicated by numbers (Tables \ref{table:sequence_ubvri} and \ref{table:sequence_griz}). The inset shows a zoom-in of the region of the SN. The size of the zoomed-in region is shown by a $16\arcsec \times 16\arcsec$ blue box.
The cyan lines show the range in slit positions during epochs when the host galaxy was included in the slit.
}
\label{fig_local_image}
\end{figure}

The observational diversity of SNe arising from massive stars ($\gtrsim$8\,$M_\odot$) is well-documented. However, not only are SNe~IIP arguably the most homogeneous subgroup, they are
also the most common type of core-collapse SN, accounting for $\sim$50\% of the volume-limited sample presented in \citet{Li_2011}. Nevertheless, it is important to bear in mind that SNe~IIP do display a range of plateau lengths, luminosities, expansion velocities, and post-plateau
decline rates. These variations can be relatively well-understood within the framework of current models \citep[e.g.][]{Kasen_2009,Dessart_2010}. 
Therefore, deviations from this observed and expected behaviour are all the more striking.
\citet{Hamuy_2003} considered a sample of 24 SNe~IIP and attempted to identify trends in physical properties such as the ejected mass, progenitor radius, and explosion energy. More recent studies are based on generally larger samples of SNe~IIP/L that are constructed using particular properties e.g. those that explore the post-peak (e.g., \citealt{Anderson_2014}, \citealt{Sanders_2015}) and pre-peak \citep{Gall_2015,Gonzalez_2015} behaviour. Many of these seem to be in broad agreement with the findings of \citet{Hamuy_2003}, in that the parameter space of physical properties occupied by SNe~IIP/L appears to be a continuous distribution, but other studies favour a clearer division between the two subtypes \citep[e.g.][]{Arcavi_2012, Faran_2014}.

In this study, we present optical observations of SN LSQ13fn, which exhibits properties characteristic of SNe~IIP, but is catapulted into a regime of parameter space that has previously remained unoccupied by the SN~IIP population by virtue of its combination of peculiarities.
We discuss several scenarios and attempt to arrive at plausible conclusions concerning both the observed
properties, and the nature of the progenitor itself.
The paper is organized into relatively self-contained sections as follows:
in \S \ref{sec:obs} we give details of the discovery, followup campaign and data reduction. In \S \ref{sec:lightcurves} we discuss the light curves. 
\S \ref{sec:spectroscopy} includes an analysis of the spectroscopy and comparisons with previous SNe. 
In \S \ref{sec:spec_analysis} we highlight and discuss the peculiar features of LSQ13fn. 
In \S \ref{sec:metallicity} we compare the spectra with radiative transfer models, and a final discussion and conclusions can be found in \S \ref{sec:discussion}.

\section{Data acquisition and reduction}
\label{sec:obs}
\subsection{Discovery, redshift, and environment}
\label{sec:discovery}

LSQ13fn was discovered by the La Silla-QUEST Variability Survey (LSQ; \citealt{Baltay_2013}) on 2013 Jan. 10.2 UT. 
It was classified by the Public ESO Spectroscopic Survey of Transient Objects (PESSTO; \citealt{Smartt_2015}) as a narrow-lined SN (type IIn), based on its apparent similarity to the prototypical SN~IIn, 1998S \citep{Sollerman_2013}.
The target was identified by PESSTO as a scientifically interesting object for follow-up and a monitoring campaign was initiated.
Our follow-up imaging and spectroscopic data revealed it to be of the SN~IIP variety (see \S \ref{sec:lightcurves}, \ref{sec:spectroscopy}). As the last non-detection of the object in LSQ images was on 2012 Dec. 23.3, 17.9 days prior to the discovery, it is not possible to precisely constrain the explosion epoch. The time of peak magnitude in the $V$ band occurred on 2013 Jan. 14.2 (MJD 56306.2). In what follows, we refer to this as day 0, and all relative epochs are in the rest frame of the SN.

\begin{figure}[t!]
\centering
\includegraphics[width=1\linewidth,angle=0]{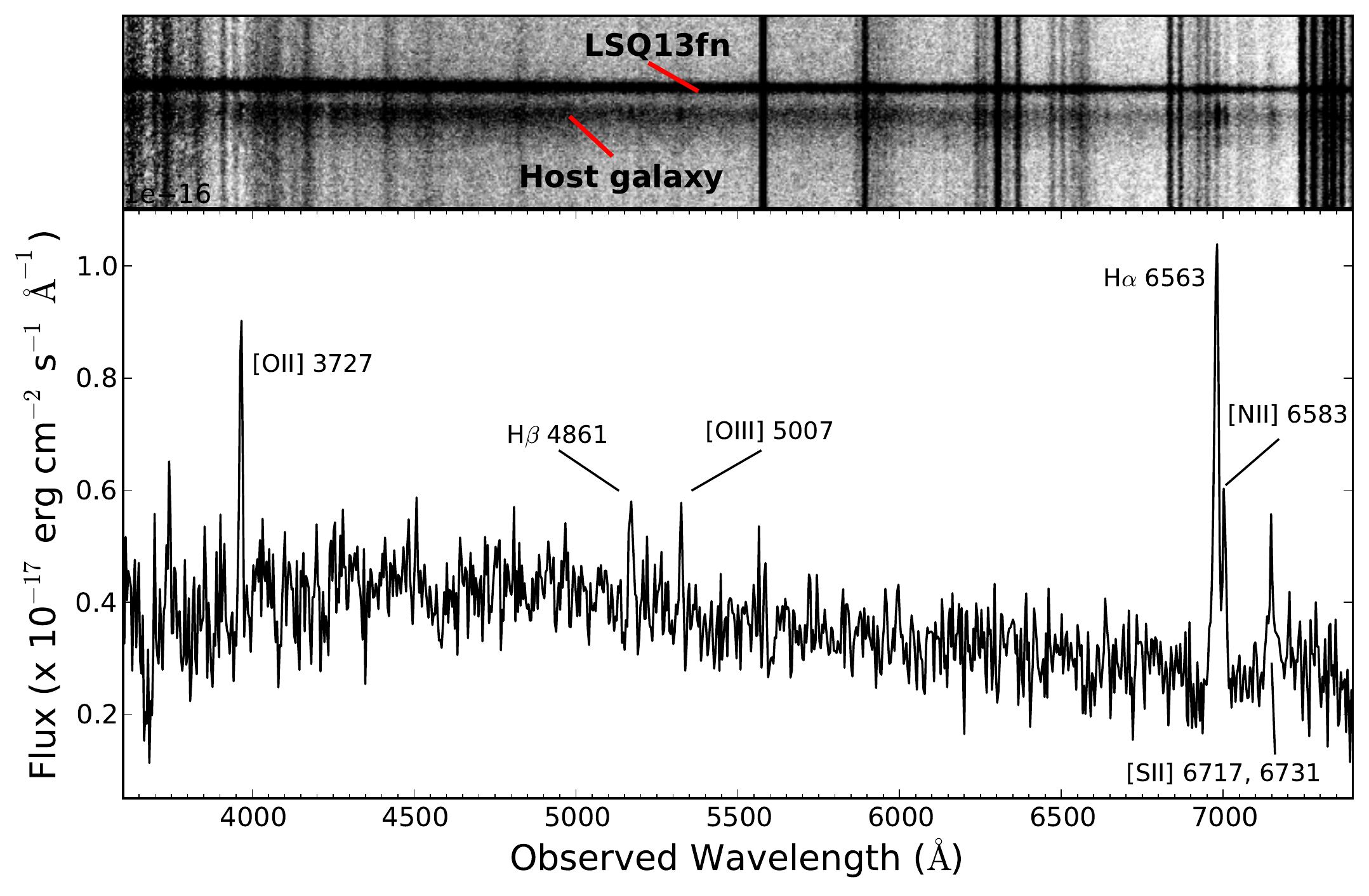}
\caption{
\textbf{Top:} 2D spectrum of LSQ13fn, taken with the NTT+EFOSC2 at 24.4\,d. The host galaxy was fortuitously also included in the slit.
\textbf{Bottom:} The extraction of the host galaxy spectrum shown in the top panel, with detected emission lines marked. The wavelength scale is aligned with the 2D spectrum. The spectrum has been corrected for Milky Way reddening only.
\label{fig_host_spec}}
\end{figure}

The edge of the presumed host galaxy (GALEXASC J115117.64-293638.0; Fig. \ref{fig_local_image}) was fortuitously included in the slit during six of the ten spectroscopic epochs.
The range in slit position angles used among the epochs was between $\sim$75$-$90$^{\circ}$ (subtended from N to E), and the galaxy spectrum was located on average $\sim$5$''$ (or at a projected distance of 6.4\,kpc) from the SN and $\sim3\farcs9$ (or 4.9\,kpc) from the galaxy nucleus. 
Several narrow emission lines were apparent in the galaxy spectrum that could easily be identified (see Fig. \ref{fig_host_spec}). 
We used these lines to calculate a redshift of $z=0.063\pm0.001$ by measuring the central wavelengths and averaging over each epoch. 
This redshift corresponds to a distance modulus of $\mu=37.1\pm0.1$ mag ($d=264$\,Mpc), with $H_{0}=75$\,km\,s$^{-1}$\,Mpc$^{-1}$, $\Omega_{M}=0.27$ and $\Omega_{\Lambda}=0.73$, and is completely consistent with that derived from the SN classification spectrum \citep[$z=0.064$;][]{Sollerman_2013}.
We estimated the oxygen abundance of the host galaxy spectra using the $N2$ calibration of \citet{pettini_pagel_2004}. 
The calibration is based on the flux ratio of emission lines that are close in wavelength: [\ion{N}{ii}]~$\lambda6583$ and H$\alpha~\lambda6563$ -- thus the sensitivity to precise flux calibration and extinction is much reduced. 
Taking the mean over all six extracted spectra results in an oxygen abundance of $12+\mathrm{log(O/H)}=8.6\pm0.2$, which is close to the solar value \citep[$Z_{\odot}=8.69$;][]{Asplund_2009}.

\subsection{Data reduction: imaging}\label{sec:imaging_reduction}

We carried out an optical photometric followup campaign using a number of different facilities: the 3.58m New Technology Telescope + EFOSC2 (NTT; $U\#640$, $B\#639$, $V\#641$, $R\#642$, $i\#705$ filters), the 2m Liverpool Telescope + RATCam (LT, $BVg'r'i'z'$ filters) and the 1.3m Small and Moderate Aperture Research Telescope System + ANDICAM (SMARTS, operated by the SMARTS Consortium; $BVRI$ filters). 
The NTT, LT, and SMARTS images were reduced using standard procedures within custom pipelines, and, when necessary, cosmic rays were removed using the {\sc lacosmic} algorithm \citep{dokkum_2001}.
We also present both pre- and early post-explosion data from LSQ. 
A log of the optical imaging is given in Tables \ref{table:photometry_ubvri} and \ref{table:photometry_griz}.

Due to the background at the location of the SN it was difficult to perform precise photometry when the source was faint and during conditions of poor seeing. 
For this reason, we gathered template images to subtract from our followup images and facilitate the measurements. 
The template images were obtained on 2014 Jan. 8 (NTT $B\#639$, $V\#641$, $R\#642$, $i\#705$ filters, also used with SMARTS data), 2014 Jan. 29 (LT $g'z'$ filters) and 2014 Feb. 8 (LT $r'i'$ filters). 
The SN was not detected in any of the images (NTT: $m_{B}>23.6$\,mag, $m_{V}>23.9$\,mag, $m_{R}>23.8$\,mag, $m_{i}>23.1$\,mag; LT: $m_{g}>22.2$\,mag, $m_{r}>22.4$\,mag, $m_{i}>22.0$\,mag, $m_{z}>20.6$\,mag).
For the LSQ data, we stacked all available pre-explosion images to create the template. The templates were aligned to each followup image using the {\sc iraf}\footnote{Image Reduction and Analysis Facility, distributed by the National Optical Astronomy Observatories, which are operated by the Association of Universities for Research in Astronomy, Inc, under contract to the National Science Foundation.} tasks {\sc geomap} and {\sc geotran}, and then convolved to and subtracted from each image using {\sc hotpants}\footnote{\url{http://www.astro.washington.edu/users/becker/hotpants.html}}. We did not obtain template images in the $U$ band.

Photometric zero points and colour terms were determined on all photometric nights using standard stars to calibrate a local sequence of stars to both the Johnson-Cousins ($UBVRI$) and SDSS\footnote{\url{www.sdss3.org}} ($griz$) systems. These stars are labelled in Fig. \ref{fig_local_image} and their magnitudes are given in Tables \ref{table:sequence_ubvri} and \ref{table:sequence_griz}, respectively. 
The LSQ images were taken using the ESO 1.0m Schmidt Telescope with a wideband filter \citep{Baltay_2013}, and we calibrated the magnitude of the SN to the $V$ band. It was not possible to apply a colour transformation because there are no available images with other filters, however it is likely that the large uncertainties of the LSQ magnitudes are greater than the differences a colour transformation would make.

Point-spread function (PSF) fitting differential photometry was carried out on all images using the custom built SNOoPY\footnote{SuperNOva PhotometrY, a package for SN photometry implemented in IRAF by E. Cappellaro; http://sngroup.oapd.inaf.it/snoopy.html} package within {\sc iraf}. 
The uncertainties of the PSF fitting were estimated by performing artificial star experiments. Artificial stars of the same magnitude as the SN were placed at several positions around the SN. The standard deviation of their magnitudes was combined in quadrature with the zero point to give the final photometric uncertainties. 

At 121 and 122\,d, the SN was not detected in the images. We determined the limiting magnitudes of the images by creating a PSF from nearby stars and adding it to the position of the SN. The magnitude was decreased until it was no longer detectable to 3$\sigma$; the faintest detection was taken to be the limiting magnitude. The same technique was applied to the pre-explosion LSQ image at $-20$\,d.

In addition to the optical photometry, a single epoch of near-infrared (NIR) photometry was obtained with the NTT + SOFI and the $JHK_{\mathrm{s}}$ filters on +23.7\,d. These images were reduced using standard techniques, and the magnitudes were determined via PSF fitting and calibrated using the 2MASS ($JHK$) catalogue \citep{Skrutskie_2006}. The resulting apparent magnitudes are $J=19.35\pm0.16$, $H=19.28\pm0.16$ and $K=19.13\pm0.46$.

\subsection{Data reduction: spectroscopy}

We obtained a series of 10 optical long-slit spectra of LSQ13fn spanning $-0.9$ to $+328.4$\,d with respect to the $V$ band maximum. 
The details of the telescopes and instrument configurations are listed in Table \ref{table:spectra_log}.
All NTT spectra were reduced using the PESSTO pipeline as described in \citet{Smartt_2015}.
Briefly, the spectroscopic frames were reduced in a standard fashion, including bias subtraction, flat field correction (which removed the fringe pattern in Gr\#16 data), and wavelength calibration via arc lamps. 
The wavelength calibration was checked by measuring the the central wavelength of the [\ion{O}{i}]~$\lambda$5577 sky emission line and corrected when necessary.
Relative flux calibration was achieved by taking spectra of spectrophotometric standard stars with an identical configuration. 
The NTT spectra were additionally corrected for telluric absorption via a model spectrum of the telluric bands, again as described in \citet{Smartt_2015}.
The WHT and VLT spectra were reduced within the {\sc iraf} environment. 
In order to account for slit-losses, we used the available photometry to interpolate to the epochs of spectroscopy and derive appropriate scaling factors to apply to the spectra. 

All reduced PESSTO EFOSC2 spectra and the SOFI imaging data are available from the ESO Science Archive Facility as Phase 3 data products. Details on how to access these data are available on the PESSTO 
website\footnote{\url{www.pessto.org}}. The reduced and astrometrically calibrated EFOSC2 images are also available from the
PESSTO website. All spectra, including the WHT and VLT data are also publicly available on WISeREP\footnote{\url{http://www.weizmann.ac.il/astrophysics/wiserep/} 
\citep{Yaron_2012}}.

\subsection{Reddening}

The Galactic reddening in the direction of LSQ13fn is $E(B-V)=0.0538$ mag \citep{Schlafly_2011}. 
A technique frequently used to estimate host galaxy extinction for SNe is measuring the equivalent width of the narrow interstellar absorption lines due to \ion{Na}{i} D \citep{Turatto_2003, Poznanski_2012}. 
Throughout the evolution of LSQ13fn, we do not detect narrow \ion{Na}{i} D lines in the spectra at the redshift of the host galaxy (see \S \ref{sec:spec_evol} and inset of Fig. \ref{fig_spectra_sequence}b), implying a low host galaxy reddening. 
We note that the relation between narrow \ion{Na}{i} D absorption and reddening has significant scatter and has been shown to be unreliable for use with low-resolution spectra \citep{Poznanski_2011}.
Other circumstantial evidence to support low host galaxy reddening comes from the location of LSQ13fn at a projected distance of $\sim$8\,kpc from the nucleus of the host galaxy, and from a consideration of our earliest spectra (Fig. \ref{fig_spectra_sequence}a) which shows a steeply rising blue continuum.
By artificially increasing the reddening and attempting blackbody fits to the spectrum, a limit of $E(B-V)\lesssim0.2$\,mag was found, assuming $R_{V}$=3.1, at which point the blackbody temperature exceeds $\sim77$\,kK.
Therefore, we treat the host galaxy reddening as negligible and correct only for Galactic reddening.

\section{Light curves and colour evolution}\label{sec:lightcurves}

\begin{figure}[t!]
\centering
\includegraphics[width=1\linewidth,angle=0]{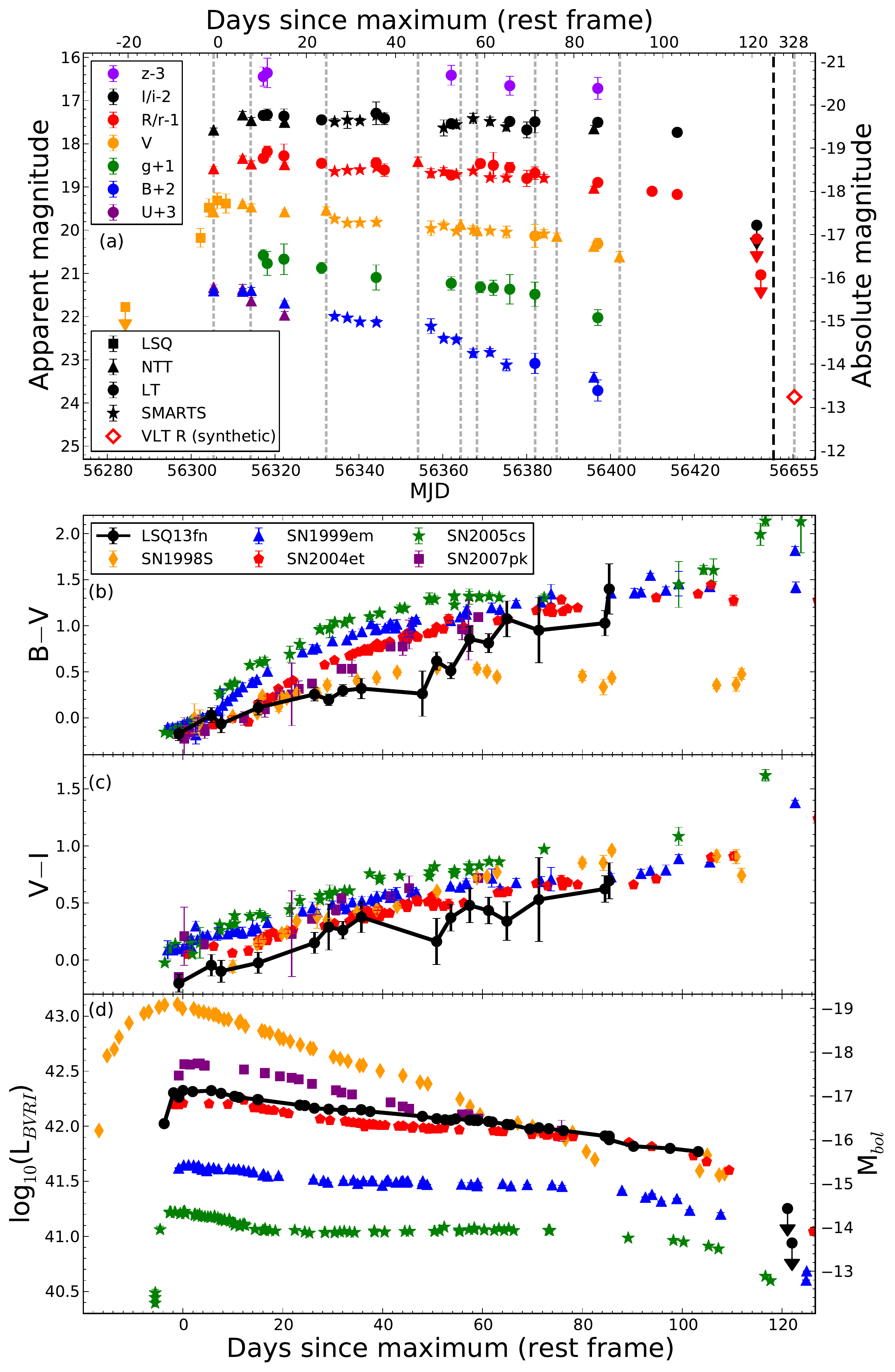}
\caption{
\textbf{Panel (a):} Lightcurves of LSQ13fn, corrected for Galactic reddening of $E(B-V)=0.0538$\,mag, using $R(V)=3.1$ and the extinction law of \citet{Cardelli_1989}. 
The absolute magnitude scale was determined by applying the distance modulus ($\mu=37.1$\,mag).
Downward pointing arrows indicate limiting magnitudes. 
Vertical grey dashed lines represent the epochs of the presented spectra of LSQ13fn, while the bold vertical dashed line marks a discontinuity in the x-axis scale.
\textbf{Panel~(b):} $B-V$ colours of LSQ13fn and comparison SNe.
\textbf{Panel~(c):} $V-I$ colours of LSQ13fn and comparison SNe. For LSQ13fn, a combination of the $V-I$ and $V-i$ colours is shown. 
\textbf{Panel~(d):} Pseudo-bolometric light curves of LSQ13fn and comparison SNe. No bolometric correction has been attempted. 
The $M_{bol}$ axis was determined assuming $M_{bol, \odot}=4.83$ mag and $L_{\odot}=3.846\times10^{33}$ erg s$^{-1}$.
\label{fig_photometry}}
\end{figure}

The photometry of LSQ13fn is given in Tables \ref{table:photometry_ubvri} and \ref{table:photometry_griz}, and the multi-band light curves are shown in Fig. \ref{fig_photometry}a. 

The $V$ band reached a peak magnitude of $19.30\pm0.18$ mag (corresponding to $M_{V}=-17.95\pm0.21$ mag, including the uncertainty of the distance modulus) at 3.8\,d after discovery. 
After the initial rise, the $V$, $R/r$, $I/i$, and $z$ bands declined slowly, at rates of $<$\,1\,mag\,100\,d$^{-1}$, while the $g$ and $B$ bands declined more rapidly, at rates of $\sim$\,1.4 and $\sim$\,2.5\,mag\,100\,d$^{-1}$, respectively. 
We measured a decline of 0.28\,mag in the $R$ band from peak to 50\,d. 
These decline rates are consistent with various classification criteria commonly used for SNe~IIP. 
For example, \citet{Patat_1994} suggested that the decline of the $B$ band should be less than 3.5 mag per 100\,d on the plateau, while \citet{Li_2011}, based on a large and homogeneous volume-limited sample including both SNe IIP and IIL, suggested the $R$ band decline should be no more than 0.5 mag during the first 50 days, both of which are satisfied by LSQ13fn.

Between 103 and 121\,d, the $r$ and $i$ bands declined by more than 1.9 and 2.1 mag respectively. 
We interpret this decline to be the end of the photospheric phase. 
The resulting duration of the plateau of LSQ13fn can be inferred to be $112\pm9$\,d.
This is within the range typically observed in SNe~IIP (e.g. Fig. 5 of \citealt{Hamuy_2003}), although somewhat longer than the average of 80--90 days \citep{Anderson_2014}.

The $B-V$ and $V-I$ colour evolution of LSQ13fn is shown in Figs. \ref{fig_photometry}b and \ref{fig_photometry}c, along with a number of SNe IIP for comparison (see Table \ref{table:SN_properties} for details and references for each SN). 
Our choice of comparison SNe is motivated simply by a combination of apparent photometric and/or spectroscopic similarities. 
The $B-V$ colour of LSQ13fn is similar to SN~2004et for the initial $\sim$20\,d.
Between $\sim$20 and 60\,d, the $B-V$ colour is marginally more blue than the comparison SNe~IIP, however LSQ13fn has a similar $B-V$ to the bluer SNe~IIP such as SN~2004et after $\sim$60\,d.
The same overall behaviour is seen in the $V-I$ colour, with LSQ13fn being marginally bluer than the comparison SNe, although the errorbars are relatively large. 

The pseudo-bolometric light curves of LSQ13fn and comparison SNe are shown in Fig. \ref{fig_photometry}d. Each light curve was computed by converting the broad-band magnitudes to flux, after correcting for reddening. 
The resulting SED was integrated over all wavelengths, assuming a zero flux level at the limits of the integration. 
Flux was then converted into luminosity according to the distance. 
For epochs during which magnitudes of a particular band were missing, a simple interpolation was performed using a low order polynomial. 
Since the coverage of the $U$, $z$, $J$, $H$, and $K$ bands of LSQ13fn are very sparse, only the $BVRI$ and $gri$ bands were used in the calculations to allow for a fair comparison between the SNe. 
The pseudo-bolometric light curve of LSQ13fn is most similar to that of SN~2004et during the plateau phase, and has a very similar shape to the other SNe IIP. 

As mentioned previously (\S \ref{sec:discovery}), the similarities in the pre-maximum spectra of LSQ13fn and SN~1998S led to a type IIn classification for LSQ13fn \citep{Sollerman_2013}.
In order to assess the presence of interaction with circumstellar material (CSM) in LSQ13fn, we show the colour evolution and pseudo-bolometric light curves of SN~1998S, a prototypical SN IIn, and SN~2007pk, a SN II with clear signs of CSM interaction at early epochs \citep{Pritchard_2012, Inserra_2013}. 
The $B-V$ colour evolution of SN~1998S is only initially similar to that of LSQ13fn, and remains bluer than all other objects shown in Fig. \ref{fig_photometry}b throughout its evolution, while the colours of SN~2007pk are similar to the other SNe~IIP. 
The pseudo-bolometric light curve of SN~1998S bears little similarity to LSQ13fn and other SNe IIP, while that of SN~2007pk is both more luminous and more rapidly declining than that of LSQ13fn.
Furthermore, as will be discussed in detail in \S \ref{sec:spectroscopy}, the spectral evolution of LSQ13fn after $\sim25$ days differs from SN~1998S and SN~2007pk. 

Thus, despite the initial classification of LSQ13fn as a type IIn, the overwhelming evidence from the light curves points towards a classification of type IIP. 
Nevertheless, it exhibits a bluer colour than other SNe~IIP up to epochs close to the end of the plateau phase.
We will return to these points in \S \ref{sec:IIP_comp}.

In principle, it is possible to estimate the mass of $^{56}$Ni synthesised in SNe by comparing the decline of the late-time light curve (after $\sim$200\,d) with that of a SN with a known $^{56}$Ni mass. 
Although our imaging of LSQ13fn does not extend to the late-time tail phase, we did obtain a spectrum at 328.4\,d (see \S \ref{sec:spectroscopy}). 
From the spectrum, we measured a synthetic magnitude of $R\sim25.0$ mag using the {\sc sms} code (Synthetic  Magnitudes from Spectra) within the S3 package (Inserra et al. in prep.). 
If we assume that the late-time light curve of LSQ13fn is powered by radioactive decay, we can estimate the $^{56}$Ni mass by comparing the luminosity with that of SN~1987A at the same epoch, assuming that the fraction of trapped $\gamma$-rays is similar.
Using a $^{56}$Ni mass of 0.073 M$_{\odot}$ for SN~1987A (weighted mean of \citealt{Arnett_1989} and \citealt{Bouchet_1991}) results in a $\mathrm{M}(^{56}\mathrm{Ni})\sim0.02\,\mathrm{M}_{\odot}$ for LSQ13fn. 
We have used the magnitude of SN~1987A at 335\,d after explosion assuming a rise-time for LSQ13fn of 7\,d (see \S \ref{sec:velocity}).
We emphasise that this value is based on only the $R$ band luminosity, rather than the bolometric luminosity, and the calculation depends on several assumptions.
Therefore it is only indicative and should be treated with caution. 
For comparison, the value is similar to that of SN~1999em (0.02 M$_{\odot}$; \citealt{elmhamdi_2003}) and greater than that of the sub-luminous SN~2005cs (0.003 M$_{\odot}$; \citealt{Pastorello_2009}), and is therefore within the range broadly expected for SNe IIP.

\section{Spectroscopy and analysis}\label{sec:spectroscopy}

In this section, we first discuss each stage of the spectroscopic evolution of LSQ13fn (Fig. \ref{fig_spectra_sequence}a), followed by an examination of various peculiarities, including comparisons with other SNe.

\subsection{Spectroscopic evolution}\label{sec:spec_evol}

\begin{figure*}[t!]
\centering
\includegraphics[width=1\linewidth,angle=0]{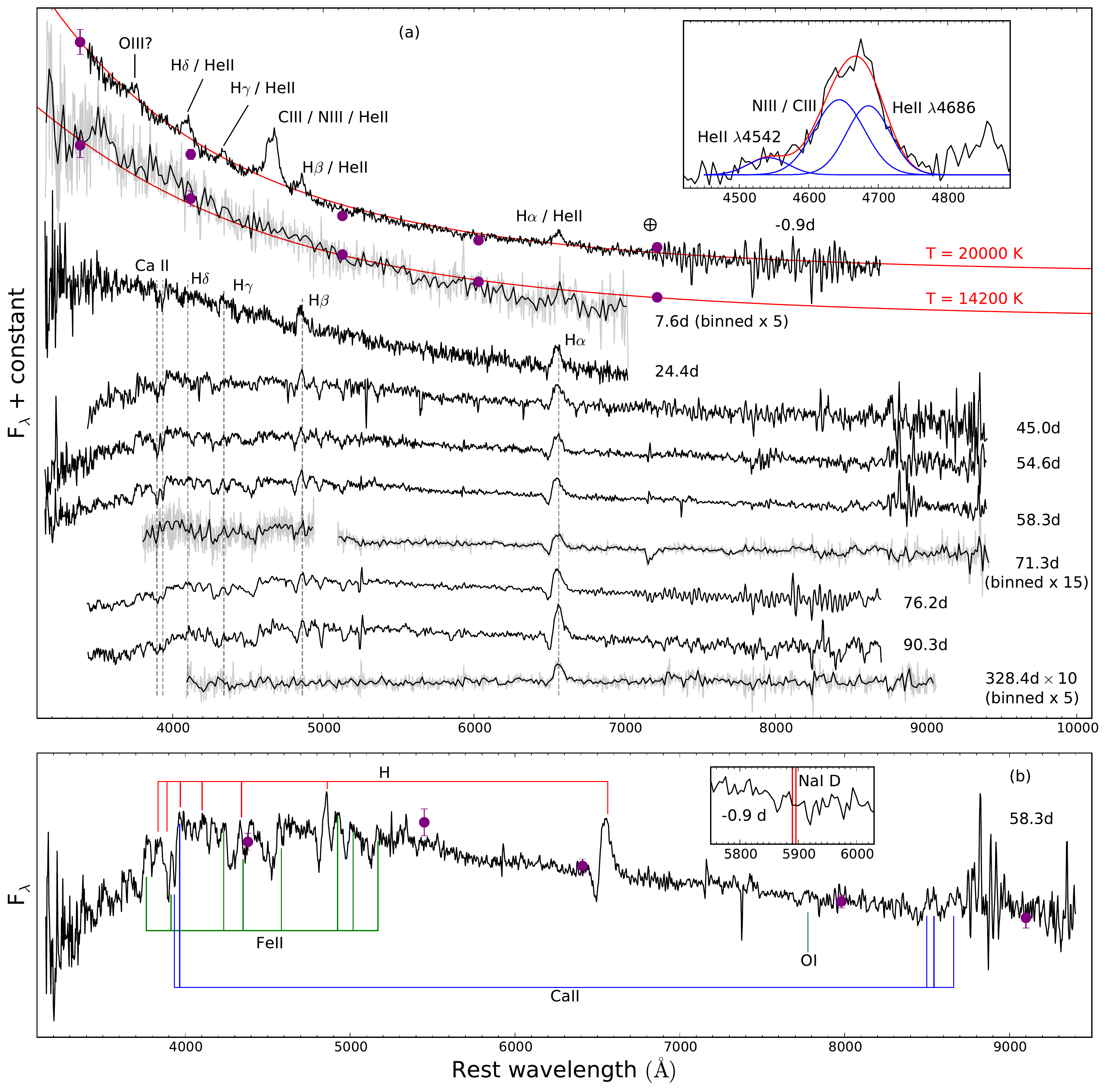}
\caption{
\textbf{Panel (a):} Sequence of optical spectra of LSQ13fn. The spectra have been corrected for reddening and redshift of the host galaxy. The epochs are in the rest frame of the SN. The 7.6\,d, 71.3\,d and 328.4\,d spectra are shown in light grey, with binned versions in black. The $\oplus$ symbol indicates the position of significant telluric absorption.
The red curves are blackbody fits to the $UBVRi$ photometry at coeval epochs (the purple points) for the $-0.9$\,d and 7.6\,d spectra.
The inset shows a fit of three Gaussian profiles to the \ion{He}{ii} / \ion{C}{iii} / \ion{N}{iii} feature of the continuum subtracted spectrum of LSQ13fn at $-0.9$\,d.
\textbf{Panel (b):} The spectrum of 58.3\,d with the strongest lines identified. The purple points show the $BVRIz$ photometry from a nearly coeval epoch. The inset shows a zoom-in of the $-$0.9\,d spectrum in the region of \ion{Na}{i} D with the rest wavelengths ($\lambda\lambda$5890, 5896) marked by vertical red lines.
\label{fig_spectra_sequence}}
\end{figure*}

The spectrum of LSQ13fn at $-0.9$\,d, obtained just two days after discovery, consists of a blue continuum with a series of strong emission features.
A least-squares fit of a blackbody function to the photometry results in a temperature of $\sim$\,20000\,K.
The rest wavelengths of some of the features are consistent with the H Balmer series at rest. 
The most striking feature in this spectrum however, lies at $\sim$4660\,{\AA}, and can be identified as a blend of \ion{N}{iii} $\lambda$4640, \ion{C}{iii} $\lambda$4648 and \ion{He}{ii} $\lambda$4542 and $\lambda$4686.
We obtain convincing fits to this feature (see inset in Fig. \ref{fig_spectra_sequence}a) using three Gaussian profiles after subtracting the continuum, fixing the centroids to 4542\,{\AA}, 4644\,{\AA}\footnote{This centroid wavelength was chosen since it is roughly between the \ion{N}{iii} $\lambda$4640 and \ion{C}{iii} $\lambda$4648 blends, which we are unable to resolve with the 17.7\,{\AA} resolution of this spectrum.} and 4686\,{\AA}, with FWHM velocities between 4000 and 5000\,\kms, allowing us to conclude that the lines are consistent with being at rest.
As an alternative to Gaussian functions, we found that a combination of Lorentzian profiles provide a reasonable fit to the feature.
However the presence of many blended lines and the insufficient S/N of the spectrum make it difficult to assign a firm preference for one or the other profile functions.
We note that the flux of the \ion{N}{iii} / \ion{C}{iii} blend is similar to that of \ion{He}{ii} $\lambda$4686, while \ion{He}{ii} $\lambda$4542 is comparatively weak.
The features with wavelengths consistent with H$\alpha$, H$\beta$, and H$\gamma$ are symmetric, are consistent with being at rest and have Gaussian FWHM velocities of $\sim$3500\,\kms.
However, the shape, strength and width of H$\delta$ (FWHM $\sim$5400\,\kms) is suggestive of a contribution by \ion{N}{iii} $\lambda\lambda$4098, 4103, lending credence to the identification of \ion{N}{iii} $\lambda$4640.
A possible identification for the emission feature seen at $\sim$3752\,{\AA} is a blend of \ion{O}{iii} $\lambda\lambda$3744, 3755, 3757, and 3760. A Gaussian fit to this feature provides a reasonable match with a FWHM $\sim$5800\,\kms.

The admittedly low signal-to-noise spectrum of LSQ13fn at 7.6\,d consists of an essentially featureless blue continuum at a temperature of $\sim$14200\,K, based on a blackbody fit to the photometry at this epoch.
The emission features -- with the possible exception of the feature at H$\alpha$ -- visible in the pre-maximum spectrum have disappeared. 
In Fig. \ref{fig_spectra_sequence}a, we show a binned version of the original spectrum, overlaid on the original data. 
The disappearance of the highly ionised \ion{N}{iii}, \ion{C}{iii}, and \ion{He}{ii} lines is likely due to the decrease in temperature by roughly a factor of two compared to the previous epoch.

By 24.4\,d, the spectrum remains blue, and the Balmer lines exhibit P Cygni profiles that are dominated by the emission component.
We note that the emission peak of the H$\alpha$ and H$\beta$ lines are blue-shifted by $\sim$600\,{\kms} from the rest velocity; while this is relatively low, it lies within the range observed in samples of SNe~II \citep[e.g.][]{Anderson_2014b}.
Weak \ion{Ca}{ii} H$\&$K absorption is also evident, but no other features are apparent in this spectrum.
Due to the presence of strong \ion{He}{ii} in the $-0.9$\,d spectrum it is inevitable that \ion{He}{i} should be present at some point.
However we do not detect \ion{He}{i} here or in the subsequent evolution, and conclude that it must have been missed or buried beneath the noise in the relatively low signal-to-noise spectra at 7.6\,d and 24.4\,d.

By the time of the mid-plateau in the light curve (45.0 to 76.2\,d spectra), the spectra of LSQ13fn have cooled further (blackbody temperature $\sim$6000\,K), becoming more depressed in the blue in comparison to the 24.4\,d spectrum. 
The spectra now consist of clear P Cygni profiles of the Balmer lines, \ion{Fe}{ii} features at $<5200\,\AA$ and \ion{Ca}{ii} H$\&$K.
We also detect weak \ion{O}{i}~$\lambda$7774 and the near-IR \ion{Ca}{ii} triplet. 
Fig. \ref{fig_spectra_sequence}b shows a close-up of our highest signal-to-noise spectrum with the most prominent lines marked. 
The spectra are similar to what is expected of SNe~IIP during the photospheric phase, supporting our classification of the SN as discussed in \S \ref{sec:lightcurves}.
However we note there are some significant differences, for example the lack of lines such as \ion{Sc}{ii}, \ion{Ba}{ii}, and \ion{Na}{i}~D between $\sim$5200$-$6500\,{\AA} which are usually observed in SNe~IIP during this phase (this is discussed further in \S \ref{sec:IIP_comp}).
By 90\,d, there has been relatively little evolution in the spectrum; however, a very weak line, most likely due to \ion{Na}{i} D, is now visible. 
This line has a similar width to the other metal lines, and is not due to interstellar absorption. 
Note that the feature at $\sim5250$\,{\AA}, visible in the 76.2\,d and 90.3\,d spectra is not real.

The final spectrum, in the nebular phase at 328\,d, is featureless with the exception of H$\alpha$ in emission which has a FWHM velocity of $\sim$3000\,{\kms}. This suggests that it originates from the SN rather than an underlying source such as an \ion{H}{ii} region. We have no explanation for the lack of both the [\ion{O}{i}] $\lambda\lambda$6300,6364 and [\ion{Ca}{ii}]
features which we would expect to be present at this epoch in spectra of SNe~IIP.

\subsection{The early-time spectra}\label{sec:early_spec}

\begin{figure*}[t!]
\centering
\includegraphics[width=1\linewidth,angle=0]{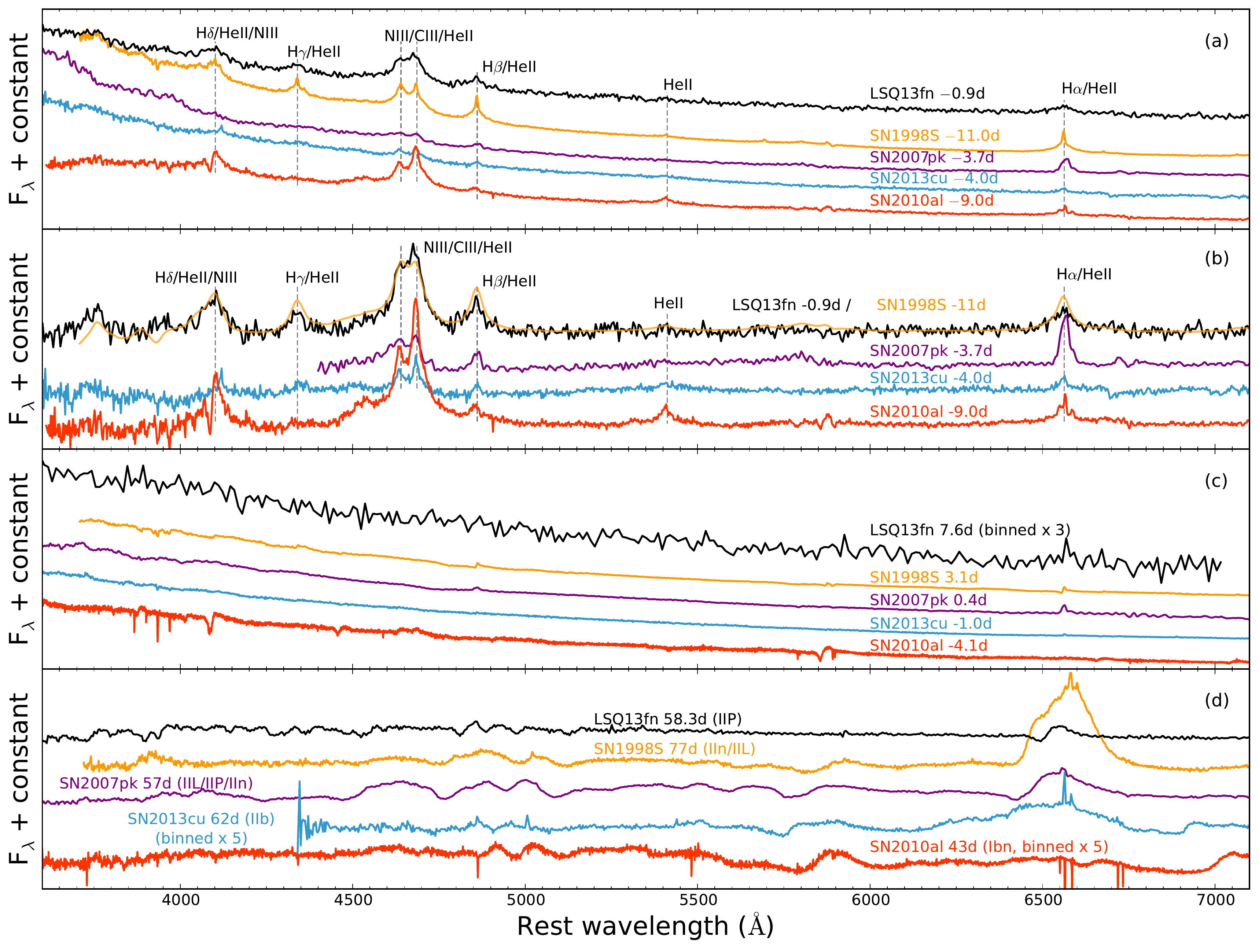}
\caption{
Comparison of spectra of LSQ13fn with some SNe that show the \ion{N}{iii}/\ion{C}{iii}/\ion{He}{ii} feature.
\citet{niemela_1985}, in the case of SN~1983K, were the first to report these lines in a SN spectrum.
All spectra have been corrected for reddening and recession velocity. 
The epochs refer to the number of days since peak magnitude.
The spectra of the comparison SNe have been scaled in flux to match the continuum level of the LSQ13fn spectra at approximately 7000\,{\AA}, and then offset by arbitrary constants, thus allowing direct comparisons between spectra.
\textbf{Panel (a):} Pre-maximum spectra of LSQ13fn and SNe that show the \ion{N}{iii}/\ion{C}{iii}/\ion{He}{ii} feature. 
\textbf{Panel (b):} Same as panel (a), but with the continuum subtracted from each spectrum. 
The spectrum of SN~1998S has been convolved with a Gaussian to match the resolution of the LSQ13fn spectrum and aligned in flux.
\textbf{Panel (c):} Spectra at epochs close to peak brightness. 
\textbf{Panel (d):} Spectra at $>$1 month after peak brightness, showing the diversity in the underlying spectral type, in contrast to early-time similarities.  
\label{fig_early_spec}}
\end{figure*}

The \ion{N}{iii}/\ion{C}{iii}/\ion{He}{ii} blend in the pre-maximum ($-0.9$\,d) spectrum of LSQ13fn has previously been observed in a handful of SNe (Fig. \ref{fig_early_spec}a)\footnote{Table \ref{table:SN_properties} lists the details and references for each SN}, although it is by no means commonplace\footnote{This may well be an observational bias due to the fact that it is challenging to obtain very early time spectroscopy, and the opposite may in fact be the case. This would not be particularly surprising given the role of mass loss in massive star evolution. }.
In order to facilitate the comparison between the SNe, low order polynomial fits to the continuum were subtracted from their spectra (Fig. \ref{fig_early_spec}b).
The $-0.9$\,d spectrum of LSQ13fn is remarkably similar to that of SN~1998S at $-11$\,d, with matching features in both strength and velocity, which underscores the original classification even though narrow lines were not evident.

\begin{figure*}[t!]
\centering
\includegraphics[width=1\linewidth,angle=0]{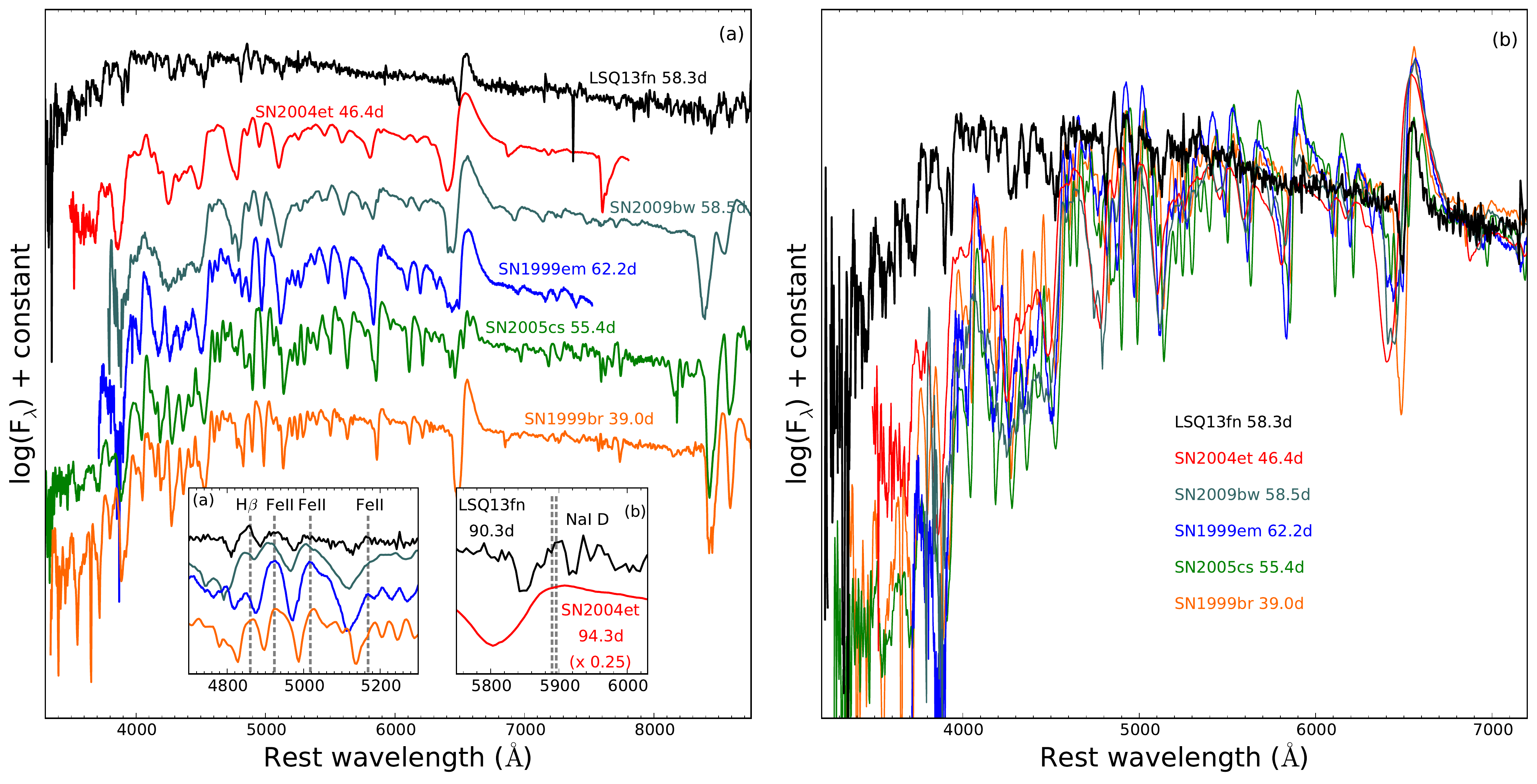}
\caption{
\textbf{Panel (a):} Comparison of spectra of LSQ13fn during the plateau phase with SNe~IIP. All spectra have been corrected for reddening and recession velocity. The epochs refer to the number of days since peak magnitude. From top to bottom, they are ordered in terms of decreasing $R$ band absolute magnitude at 50\,d. Inset (a) shows a zoom-in of SN~LSQ13fn, SN~1999br, SN~1999em, and SN~2009bw. The rest wavelengths of the strongest \ion{Fe}{ii} lines and H$\beta$ are indicated by the vertical dashed lines. Inset (b) shows a zoom-in of the weak \ion{Na}{i} D line detected in the 90.3\,d spectrum of LSQ13fn. A spectrum of SN~2004et at a similar epoch is included for comparison, which has been scaled. 
\textbf{Panel (b):} The same spectra as panel (a), with the flux aligned at $\sim7000$\,{\AA}.
\label{fig_plateau_spec}}
\end{figure*}

Fig. \ref{fig_early_spec}c shows the subsequent evolution of the SNe shown in Fig. \ref{fig_early_spec}a on timescales of about a week from the earliest epochs. 
All objects show a blue, nearly featureless continuum. 
As shown in Fig. \ref{fig_early_spec}d, the subsequent evolution of these SNe is far from homogeneous, as expected from their respective classifications that include both hydrogen-rich and hydrogen-poor core-collapse SN types. 

Given that all massive stars are likely to have some CSM in their vicinity at the time of explosion, we suggest that a likely scenario is one where the high temperatures required to generate the high ionisation lines (\ion{N}{iii}, \ion{C}{iii}, and \ion{He}{ii}) may be caused by a large flux of hard photons produced either during the UV flash after shock breakout, or interaction of the SN ejecta with a dense CSM of variable location and extent. 
As demonstrated recently by \citet{Groh_2014} for the case of SN~2013cu (type IIb), the high ionisation features arise from the ionisation of the CSM. Thus, we may expect to see such features in spectra taken soon after explosion regardless of (core-collapse) SN type.
The subsequent fading of these features is almost certainly due to cooling as the SN expands and over-runs the CSM, allowing the ejecta spectrum to finally emerge at later epochs (Fig. \ref{fig_early_spec}). 
For the specific cases of LSQ13fn and SN~1998S, the similarity in the earliest spectra does not extend to beyond the earliest spectroscopic epochs. 
Indeed, the spectroscopic evolution of SN~1998S shows signatures of ongoing CSM interaction throughout much of its evolution \citep{fassia_2001,Pozzo_2004}.

As an aside, we call into question the identification of narrow features in the earliest spectra of SN~1998S, SN~2010al, and SN~2013cu as being exclusively due to hydrogen. 
Motivated by the above suggestion regarding the influence of the rapidly evolving temperature on the appearance of the spectrum, \ion{He}{ii} transitions to the fourth energy level are a likely contribution to the narrow emission features -- usually attributed to the Balmer series. 
Indeed, this identification is supported by the clear detection of a feature at $\lambda$5412 that is due to \ion{He}{ii}, and belongs to the same series. 
This feature is not apparent in the LSQ13fn spectrum which may at least partially be attributed to the low signal-to-noise. 
A fake feature with the strength and shape matching that of SN~1998S was added to the spectrum of LSQ13fn, but was not recovered at the 3\,$\sigma$ level, meaning that we cannot rule out that the feature was present in LSQ13fn but buried beneath the noise. 
However, this spectrum was taken at an epoch that is several days later than the comparison objects, and may therefore be further along in its temperature evolution. 
Last, but not least, models of the spectra of SN~2013cu (Fig. 2 in \citealt{Groh_2014}, also see \citealt{Grafener_2015}) and SN~1998S \citep[Fig. 6 in][]{Shivvers_2014} bolster our suggestion that the early-time narrow lines are due to \ion{He}{ii} in combination with hydrogen.

\subsection{Comparison with SNe IIP during the plateau}\label{sec:IIP_comp}

\begin{figure*}[!t]
\centering
\includegraphics[width=1\linewidth,angle=0]{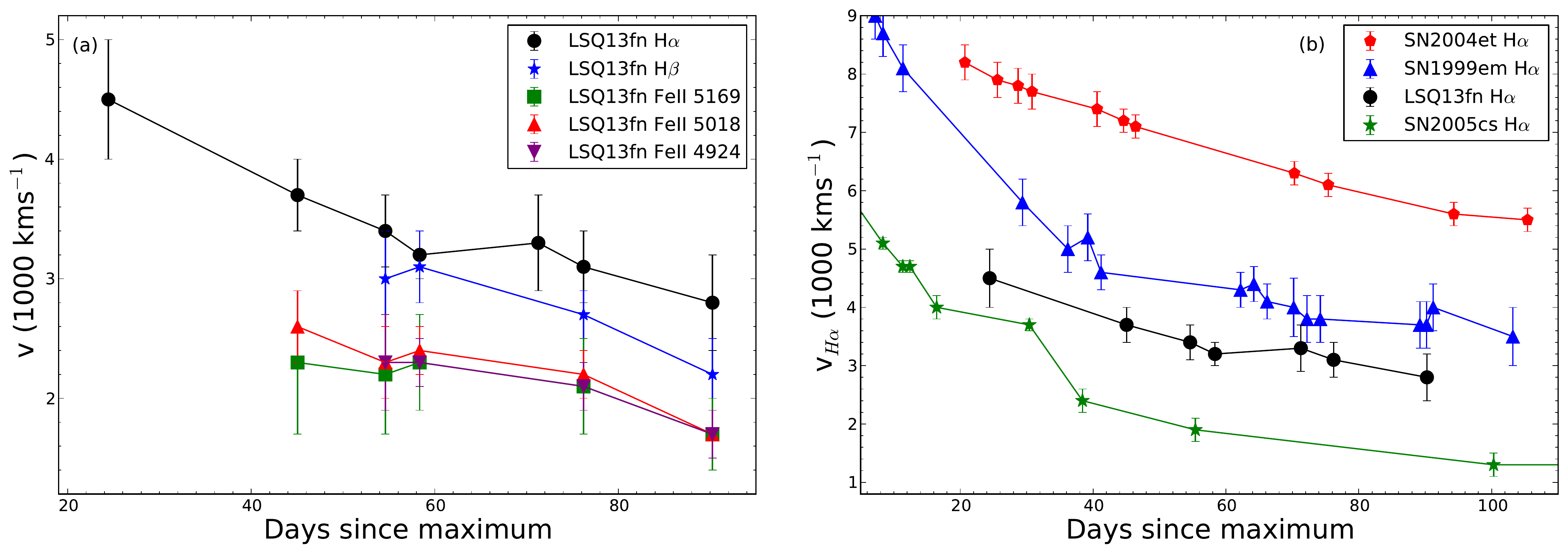}
\caption{
\textbf{Panel (a):} Comparison of the expansion velocities of various lines of LSQ13fn as measured from their P Cygni absorption minima.
The uncertainties were estimated by repeating the measurement of the absorption minimum several times across slightly different wavelength ranges.
\textbf{Panel (b):} Comparison of the expansion velocities of LSQ13fn and other SNe~II as measured from the P Cygni absorption minima of H$\alpha$.
}
\label{fig_13fn_lines_velocity}
\end{figure*}

Although LSQ13fn is consistent with being classified as a SN~IIP given its light curve and spectroscopic evolution, it does exhibit a number of peculiarities that are at odds with the garden variety of SNe~IIP. 
The high ionisation features and early-time colour evolution were discussed previously; here we focus on comparisons of the 45 to 90\,d spectra with other SNe~IIP.

The highest S/N spectrum of LSQ13fn during this phase is shown in Fig. \ref{fig_plateau_spec}a along with a selection of SNe~IIP. 
The comparison SNe span a range of plateau brightnesses, with $R$ band absolute magnitudes at 50 days after peak between $-$14.2\,mag (SN~1999br) and $-$17.5\,mag (LSQ13fn and SN~2004et). 

Compared to all other objects, there is a striking lack of strong absorption in the spectrum of LSQ13fn in the $\sim$5200$-$6500\,{\AA} range; many strong lines due to \ion{Na}{i} D, \ion{Ba}{ii}, \ion{Sc}{ii}, and \ion{Fe}{ii} are clearly seen in other SNe~IIP. 
This weakeness of features in LSQ13fn persists throughout our spectroscopic coverage, except for the 90.3\,d spectrum, in which a weak line consistent with \ion{Na}{i} D is detected (see inset (b) of Fig. \ref{fig_plateau_spec}a).
All of the comparison SNe shown in Fig. \ref{fig_plateau_spec}a show a strong \ion{Na}{i} D feature comparable in strength to the metal lines at wavelengths bluer than $\lesssim5200$\,{\AA}, and therefore the non-detection in LSQ13fn is very unlikely to be due to the S/N. 

Furthermore, although the line profiles of the features present in LSQ13fn appear to be similar to those in the comparison SNe, their strengths are significantly weaker. 
This is evident particularly in the metal lines bluewards of $\sim$5200\,{\AA}. 
Inset (a) of Fig. \ref{fig_plateau_spec}a shows a zoom-in of the spectra of LSQ13fn along with SN~1999br, SN~1999em, and SN~2009bw between 4750$-$5250\,\AA.

Fig. \ref{fig_plateau_spec}b shows the same spectra as the left-hand panel, but aligned in flux at H$\alpha$, allowing for an easier inter-comparison of the overall shape of the spectra. 
Bluewards of $\sim$5500\,{\AA} there appears to be greater flux in the continuum of LSQ13fn in comparison to the other SNe. 
In other words, the continuum of LSQ13fn is not suppressed in the blue to the same extent as the comparison SNe. 
This extra flux in the blue part of the spectra of LSQ13fn is consistent with the $B-V$ colour evolution (discussed in \S \ref{sec:lightcurves} and shown in Fig. \ref{fig_photometry}b) and the temperatures according to blackbody fits of $\sim$6000\,K, which is around 1000\,K hotter than typical of SNe~IIP at these epochs.
We finally note that the flux of H$\alpha$ in emission in LSQ13fn appears to be weak in comparison to the SNe shown in Fig. \ref{fig_plateau_spec}.

\subsection{Expansion velocity}\label{sec:velocity}

The expansion velocity of the ejecta of SNe can provide information on the energetics of the explosion.
The evolution of the H$\alpha$, H$\beta$, \ion{Fe}{ii} 4924\,{\AA}, \ion{Fe}{ii} 5018\,{\AA} and \ion{Fe}{ii} 5169\,{\AA} velocities of LSQ13fn -- as measured from the minimum of the P Cygni trough -- are shown in Fig. \ref{fig_13fn_lines_velocity}a.
There is a clear decline in the expansion velocities of H$\alpha$ and H$\beta$ with time. 
It is generally accepted that weak lines with low optical depths such as \ion{Sc}{ii} $\lambda$6246 and \ion{Fe}{ii} $\lambda\lambda$5169, 5018, 5276 are better trackers of the photospheric velocity than the Balmer lines \citep{hamuy_2001}. 
However, the evolution of the \ion{Fe}{ii} lines in Fig. \ref{fig_13fn_lines_velocity}a appears to be flat, which is perhaps due in part to the lack of a measurement at epochs earlier than 45.0\,d and the large uncertainties.
For these reasons we have chosen to use H$\alpha$ as a means of comparing the expansion velocity of LSQ13fn with other SNe IIP, as shown in Fig. \ref{fig_13fn_lines_velocity}b.
The H$\alpha$ velocities of LSQ13fn are intermediate between those of SN~2005cs and SN~1999em, and significantly lower than those of SN~2004et, in spite of the close match in plateau luminosity with the latter (Fig. \ref{fig_photometry}d).
This is somewhat surprising given the expectation of a positive correlation between the expansion velocity and plateau luminosity as has been demonstrated for SNe~IIP \citep{Hamuy_2002} i.e., the plateau luminosity of LSQ13fn is too bright considering the photospheric velocity.

\begin{figure}[!t]
\centering
\includegraphics[width=1\linewidth,angle=0]{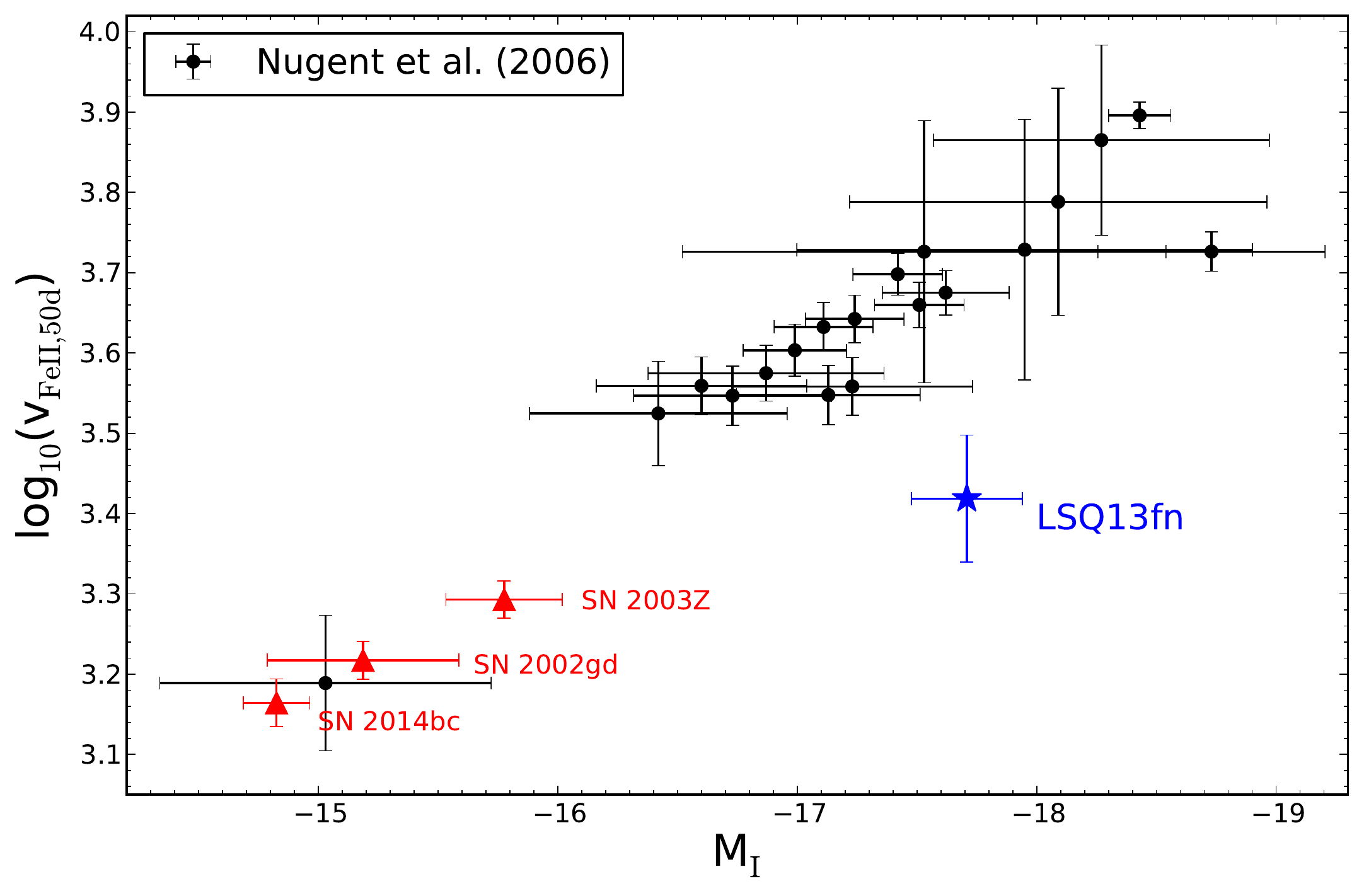}
\caption{
Velocity of the absorption minimum of the \ion{Fe}{ii} $\lambda$5169 line of LSQ13fn (blue star) against the absolute magnitude in the $I$ band, both measured at 50 days. 
The black circles are a sample of SNe IIP taken from \citet{Nugent_2006}, while the red triangles show SN~2002gd and SN~2003Z from \citet{Spiro_2014}, and SN~2014bc from \citet{Polshaw_2015}.
The velocities of SN~2002gd and SN~2003Z were measured from spectra downloaded from WISeREP and extrapolated to 50\,d via equation 2 of \citet{Nugent_2006}.
}
\label{fig_vel_mag}
\end{figure}

In order to explore this further, we apply the standardised candle relation to LSQ13fn, which is used to infer distances by exploiting the correlation between the expansion velocity of \ion{Fe}{ii} and the plateau luminosity at a fiducial epoch of 50\,d post-explosion \citep[e.g.][]{Hamuy_2002,Nugent_2006,Poznanski_2009}. 
From the \ion{Fe}{ii} $\lambda$5169 line in the 58.3\,d spectrum, we measure a velocity of $v_{\ion{Fe}{ii}}=2200\pm400$\,\kms. 
Using equation 2 of \citet{Nugent_2006} to extrapolate to 50\,d after explosion, we obtain $v_{\ion{Fe}{ii},50d}=2620\pm480$\,\kms. 
We have assumed a rise-time of 7\,d to peak, which appears to be typical for SNe~IIP from the weighted average of the rise-times with explosion epochs constrained to within $\pm$1\,d \citep{Gall_2015, Gonzalez_2015}.
Although the explosion epoch of LSQ13fn is not tightly constrained (\S\,\ref{sec:discovery}) this uncertainty is significantly smaller than that of the velocity measurement itself. 
The velocity and absolute magnitude in the $I$ band of LSQ13fn at $\sim$50\,d are shown in Fig. \ref{fig_vel_mag} along with the sample of \citet{Nugent_2006}, taken from their Table 4.

It is clear that LSQ13fn is distinct from the expected correlation. 
We note that in the unlikely event that the extinction has been underestimated, LSQ13fn would be pushed even further from the correlation. 
This means that the standardised candle method would yield an incorrect distance estimate to LSQ13fn: using equation 2 in \citet{Hamuy_2002} and assuming $H_{0}=75$\,\kms\,Mpc$^{-1}$ we find $\mu=35.2\pm0.6$\,mag, which is significantly different to the redshift-based distance modulus of $\mu=37.1\pm0.1$\,mag where the redshift was obtained from the host galaxy spectrum (\S \ref{sec:obs}).
One of the main advantages of the standardised candle method as applied to SNe~IIP is that the data requirements are relatively modest with 1 to 2 spectra encompassing the mid-point of the plateau, and photometry sufficient to ascertain that the SN is indeed of type IIP. 
However, in the absence of multi-band light curves and spectra of reasonable S/N, and no independent redshift estimates for the host galaxy -- which is often the case in studies of intermediate to high-redshift SNe -- one might have no evidence available to suggest that objects such as LSQ13fn would deviate from the standardisation. 
Although LSQ13fn may be the first object of its kind to display the above characteristics in the local Universe, it is unclear what the incidence of such SNe might be at higher redshifts
(e.g. $z\gtrsim0.1$); indeed its relative brightness would almost certainly ensure its inclusion in searches for SNe~IIP (Gall et al. 2015, in prep.).

\section{What makes LSQ13fn peculiar?}\label{sec:spec_analysis}

In the previous sections we have highlighted the following aspects of LSQ13fn that make it peculiar compared to typical SNe~IIP: 
An excess of flux bluewards of $\sim$5500\,{\AA}; a persistent lack of spectral features between $\sim$5200$-$6500\,{\AA}; weaker spectral lines than other SNe~IIP during the plateau phase; expansion velocities that are lower than expected given its absolute plateau brightness. In this section we consider two scenarios that may explain some of these peculiarities.

\subsection{CSM interaction}\label{sec:CSM_discussion}

The first possibility we consider is interaction between the ejecta and CSM. 
The presence of CSM in close proximity to the progenitor at the time of explosion could simultaneously explain the presence of the high ionisation features in the pre-maximum spectrum and the early-time excess in the bluest spectral regions, caused by the conversion of kinetic to thermal energy.
This boost of thermal energy could then cause the temperature to remain higher for a longer time, leading to the peculiar ionisation evolution of the ejecta during the photospheric phase.
Furthermore, the weakness of spectral features could be attributed to veiling as a result of the additional continuum generated from the ejecta-CSM interaction.
That LSQ13fn is too bright for its corresponding velocity could then be explained by a combination of the ejecta being slowed down by the CSM, while its brightness would be enhanced by the additional flux generated from the interaction. 

Although this scenario may seem appealing at first sight, there are a number of discrepancies that arise upon further scrutiny.
In order to explain the deviation from the velocity-luminosity relation, a boost of $\sim$1.5$-$2\,mag coming from the ejecta-CSM interaction would be necessary.
Furthermore a CSM mass comparable to the ejecta mass would be required to slow the ejecta and satisfy momentum conservation, which is predicted by hydrodynamical and radiative transfer models to be $\gtrsim$\,5$-$10\,M$_{\odot}$ for typical SNe~IIP \citep[e.g.][]{Kasen_2009,Bersten_2011,Dessart_2013}.
However, at no point in its evolution do we see narrow lines in the spectra indicative of dense material. 
This statement holds even after accounting for the modest spectral resolution of our data (Fig. \ref{fig_early_spec}b).
One may then be tempted to invoke CSM material of somewhat lower density (say, $\lesssim10^{7}$cm$^{-3}$) which would preclude the formation of narrow lines, while still providing an additional source of continuum through continued interaction with the ejecta. 
For such a scenario, a finely-tuned combination of CSM mass and velocity would be required to slow the ejecta velocities to the observed values, and would almost certainly not maintain the plateau-shaped light curves that are observed, and commensurate with the garden variety of SNe~IIP (Fig. \ref{fig_photometry}).

In order to test the idea that the weak spectral features are due to an additional source of continuum radiation, we added a blackbody to a spectrum of SN~2005cs tweaking the chosen temperature such that the H$\alpha$ line strength roughly matched that of LSQ13fn, which was found to be the case at 7000\,K. 
This procedure crudely reproduced the broad overall shape of the 58\,d LSQ13fn spectrum. 
However, the features between 5200$-$6500\,{\AA} (i.e. \ion{Na}{i}, \ion{Ba}{ii}, \ion{Sc}{ii}) were still clearly visible, while the features blueward of $\sim$4500\,{\AA} were now even weaker than those of LSQ13fn. 
Higher temperatures would make an already blue spectrum bluer still, while lower temperatures would result in lines that are too strong c.f. LSQ13fn.
We also note that all other SNe shown in Fig. \ref{fig_plateau_spec}a show spectral features in the 5200$-$6500\,{\AA} region that are comparable in strength to the metal lines at wavelengths bluer than $\lesssim5200$\,{\AA}. 
However, in LSQ13fn, we clearly detect lines blueward of $5200$\,{\AA} but find the 5200$-$6500\,{\AA} region to be virtually devoid of spectral features. 
Thus explaining the strength of all the features of the spectra as due to veiling from ejecta-CSM interaction would imply preferential suppression of spectral lines in adjacent wavelength regions, which is unphysical.

Therefore we deem that ejecta-CSM interaction was significant at the earliest epochs, and that the CSM was rapidly overrun by the ejecta.
However the conversion of kinetic to thermal energy from the early interaction appears to be able to only partly explain evolution of the spectra and the discrepancy between the luminosity and expansion velocity, which leads us to consider an alternative scenario.

\begin{figure*}[!t]
\centering
\includegraphics[width=1\linewidth,angle=0]{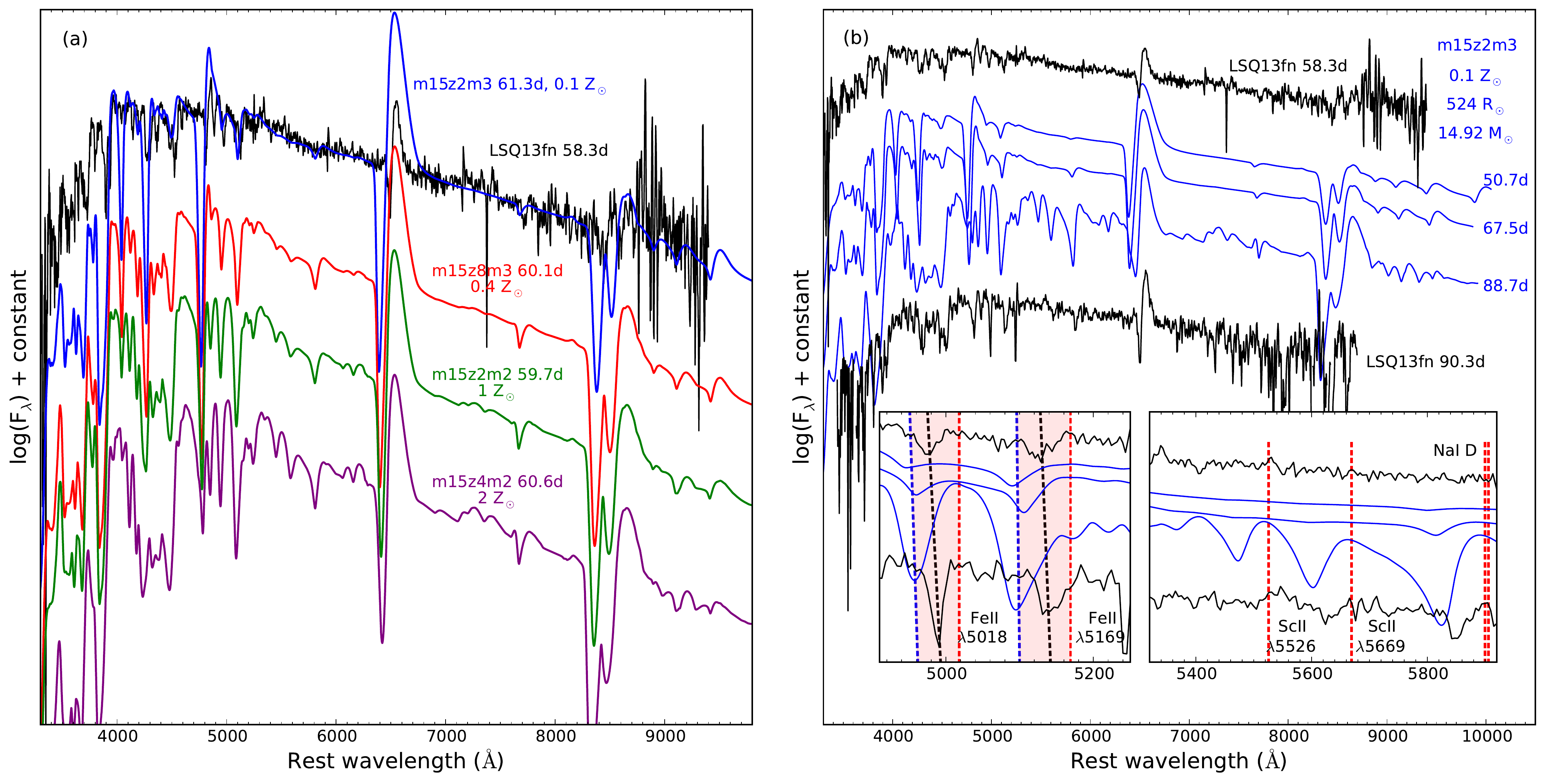}
\caption{
\textbf{Panel (a):} Comparison of the spectra of LSQ13fn with the models of \citet{Dessart_2013} of various metallicity. 
Note that the epochs of the models refer to the number of days after explosion, whereas the epochs of LSQ13fn refer to the number of days after peak magnitude in the rest frame of the SN.
\textbf{Panel (b):} Comparison of the spectra of LSQ13fn with a time-sequence of the lowest metallicity model of \citet{Dessart_2013}. The left-hand inset shows a zoom-in of the region of \ion{Fe}{ii} $\lambda\lambda$5018, 5169. The dashed red lines indicate the rest wavelengths of the two lines. The dashed black and blue lines (for LSQ13fn and the models respectively) and the shaded regions are designed to guide the eye to the absorption minima caused by the two \ion{Fe}{ii} lines. The right-hand inset shows a zoom-in of the region encompassing \ion{Sc}{ii} $\lambda\lambda$5526, 5669 and \ion{Na}{i} D.
\label{fig_dessart_spec}}
\end{figure*}
 
\subsection{Low metallicity}\label{sec:peculiar_metallicity}

An alternative explanation for the aforementioned peculiarities of LSQ13fn and their evolution in the post-24.4\,d spectra is that the SN ejecta were metal poor.
The lack of metal lines detected between 5200$-$6500\,{\AA} and their overall weakness may be indicative of a low metal abundance. 
In addition, the spectra of SNe are very sensitive to the composition of the photosphere in the ultraviolet - and by extension the blue part of the optical range - due to line blanketing by metals. 
The persistent blueness of the LSQ13fn spectra (Fig. \ref{fig_plateau_spec}b) suggests that it is less affected by line blanketing compared to the other SNe~IIP.
However, low metallicity alone cannot account for the slow expansion velocities and luminous plateau.
We therefore suggest that a combination of early-time CSM interaction and low metallicity in the SN ejecta can explain the majority of the peculiarities in LSQ13fn.
In the following section, we explore this idea further by comparing the spectra of LSQ13fn with synthetic spectra of SNe~IIP generated for a range of metallicities.

As discussed in \S \ref{sec:discovery}, the metallicity of the host galaxy, measured at projected distances of $\sim$6.4\,kpc from LSQ13fn and $\sim$4.9\,kpc from the galaxy nucleus, was estimated to be $12+\mathrm{log(O/H)}=8.6\pm0.2$.
Given the type SA\footnote{a rough classification based on our images.} morphology of the host galaxy (Fig. \ref{fig_local_image}), it is not surprising to find a value for the metallicity that is close to solar\footnote{We considered the possibility of a faint and possibly diffuse companion galaxy at the same redshift. Our limits based on template images and are given in \S\ref{sec:imaging_reduction}, but deeper imaging would be required.}. 
Considering the radially decreasing metallicity gradients of galaxies \citep[e.g.,][]{Henry_1999} one would expect a lower metallicity at the location of LSQ13fn, although it is worth noting that in the majority of galaxies the metallicity gradient appears to flatten towards the outskirts \citep{Sanchez_2014}.
Indeed the literature abounds with studies that emphasize the discrepancies between the host central regions and SN site metallicities (e.g. \citealt{Modjaz_2011}). 
All of the above notwithstanding, the progenitor metallicity itself is likely to be the dominant factor insofar as metallicity effects play a role in SN spectra, so we investigate this further in the next section.

\section{Comparison with synthetic spectra}\label{sec:metallicity}

In this section we investigate the possibility that the ejecta of LSQ13fn was metal-poor, based on the evidence discussed in \S \ref{sec:spec_analysis}, by comparing with synthetic spectra generated from radiative transfer models of SNe~IIP and their progenitors.
We also investigate the effects of the choice of initial radius.

\citet{Dessart_2013} and \citet{Dessart_2014} (hereafter D13 and D14, respectively) presented a grid of models to investigate the effects on SN~IIP light curves and spectra as a function of various progenitor parameters. 
The models of interest here follow the evolution of main-sequence stars of initial mass 15 M$_{\odot}$ and a variety of initial metallicities, while all other initial parameters are kept constant.
The evolution of the stars are computed consistently with a stellar evolution code, and therefore the initial metallicity influences the nuclear burning in a complex way, along with the opacity and mass-loss.
Consequently, although the models initially only differ by metallicity, they result in SN progenitors which differ by mass, radius, and envelope composition, reflecting the initial choices in metallicity.

\begin{figure}[t]
\centering
\includegraphics[width=1\linewidth,angle=0]{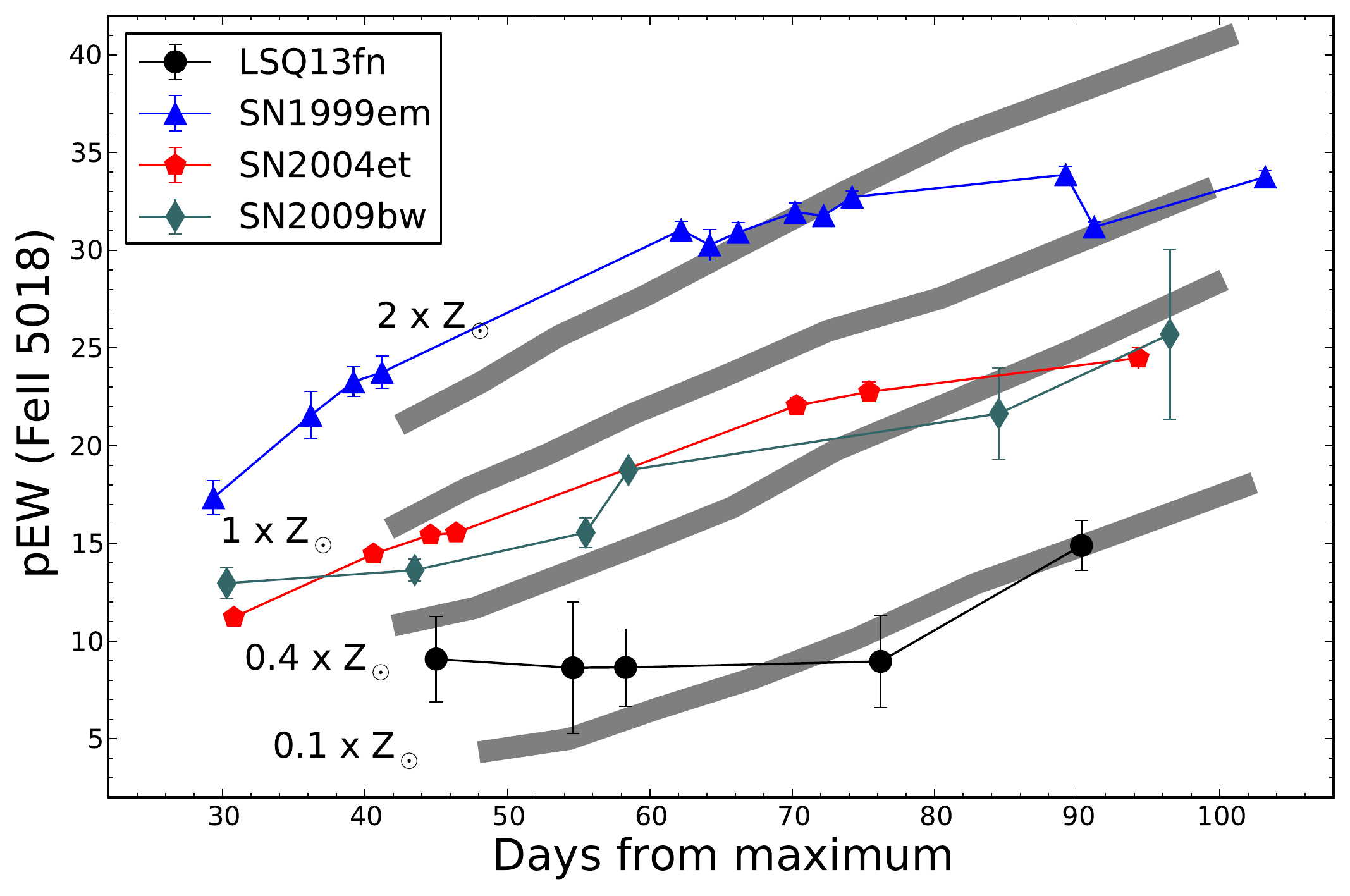}
\caption{Measurements of the pEW of the \ion{Fe}{ii} $\lambda$5018 line in the spectra of LSQ13fn and comparison SNe~IIP. 
The thick grey lines show measurements from the model spectra of \citet{Dessart_2014}. 
The epochs of the model spectra, originally referring to the number of days after explosion, have been shifted to the number of days after maximum light assuming a rise time of 7\,d (see \S \ref{sec:velocity} for a justification of this choice).
We note that oxygen abundance of nearby \ion{H}{ii} regions have previously been measured for the comparison SNe shown in the figure: $12+\mathrm{log(O/H)}\sim8.6$, 8.3, and 8.66 for SN~1999em \citep{Smartt_2009}, SN~2004et \citep{Smartt_2009}, and SN~2009bw \citep{Inserra_2012} respectively.
Assuming a Solar abundance of $12+\mathrm{log(O/H)}=8.69$ \citep{Asplund_2009}, these correspond to approximately 0.8\,Z$_{\odot}$, 0.4\,Z$_{\odot}$, and 0.9\,Z$_{\odot}$ respectively for SN~1999em, SN~2004et, and SN~2009bw.
\label{fig_pEW}}
\end{figure}

D14 specifically focuses on the observable effects of varying the metallicity in the $0.1-2\,Z_\odot$ range during the photospheric phase as this material is only weakly affected by nuclear burning for elements more massive than \ion{O}.
Therefore, the intermediate-mass and \ion{Fe}-group elements observed in the SN spectra during this phase approximately reflect the composition of the material from which the progenitor star was formed.
They found that the line strengths of lines such as \ion{O}{i} $\lambda$7774, \ion{Na}{i} D, the \ion{Ca}{ii} triplet and \ion{Fe}{ii} to be particularly strongly correlated by variations in the metallicity.

In Fig. \ref{fig_dessart_spec}a we compare the spectra of LSQ13fn with the models 
of varying metallicity presented in D13. 
We stress that the models are not tailored to parameters that are necessarily appropriate for LSQ13fn. 
Nevertheless, we find that LSQ13fn bears the greatest resemblance to the lowest metallicity model of D13 (m15z2m3, with $0.1\,Z_{\odot}$), which we show a time series of in Fig. \ref{fig_dessart_spec}b. 
At 50.7 and 67.5\,d, the m15z2m3 model shows very weak lines between $\sim$5200-6500\,{\AA}, just as in LSQ13fn. 
The \ion{Fe}{ii} and \ion{Sc}{ii} lines are weak, as shown in the insets of Fig. \ref{fig_dessart_spec}b. 
Between $\sim$4000$-$5000\,{\AA} the lowest metallicity model shows features that are stronger; however, the correspondence between the features present e.g. comparing the 90\,d observed spectrum and the 89\,d model is reassuring.
As expected, the model also does not show as large an amount of line blanketing as the higher metallicity models.
Although the overall shape of the SED is very similar to the spectrum of LSQ13fn, we note that at wavelengths $\lesssim$4000\,{\AA} the model has less flux than LSQ13fn.
By $\sim$88.7\,d the strength of the lines in the model, particularly \ion{Sc}{ii} $\lambda\lambda$5526, 5669, and \ion{Na}{i} have significantly increased in strength. 
While the \ion{Na}{i} line is detected in the 90.3\,d spectrum of LSQ13fn, it is comparatively very weak, and \ion{Sc}{ii} and $\lambda\lambda$5526, 5669 are not detected at all. 
The Balmer lines are stronger in the models than in the spectra of LSQ13fn at all epochs. 
We speculate that perhaps LSQ13fn evolved more slowly than the model, potentially due to `over-ionisation' from the residual thermal energy from the early time CSM interaction, or that even lower metallicity models might be better matched. 

\begin{figure}[t]
\centering
\includegraphics[width=1\linewidth,angle=0]{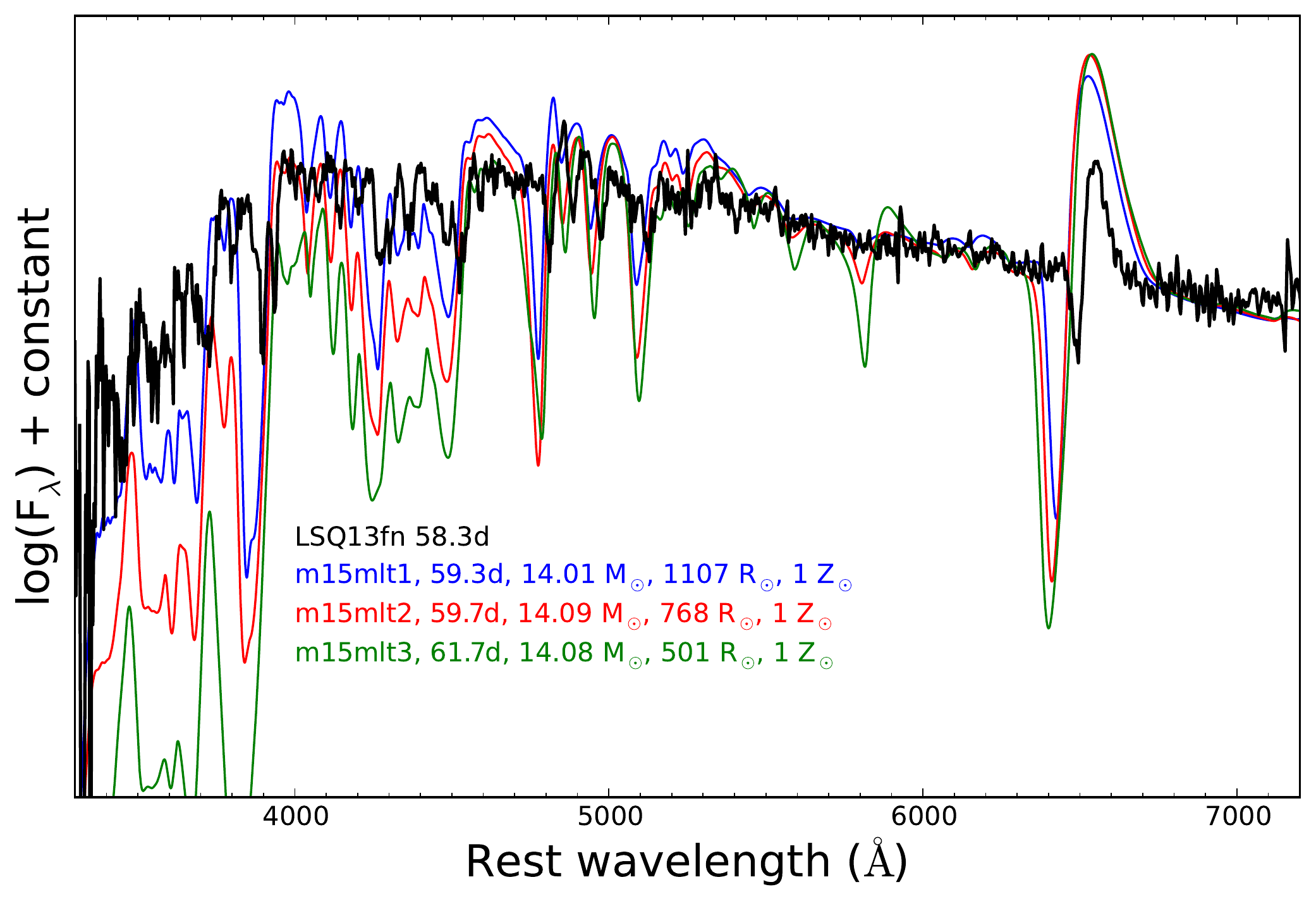}
\caption{
Comparison of the 58.3\,d spectrum LSQ13fn with three models of \citet{Dessart_2013} which have varying radii but otherwise approximately identical initial conditions.
Note that the epochs of the models refer to the number of days after explosion, whereas the epochs of LSQ13fn refer to the number of days after peak magnitude. The spectrum LSQ13fn is shown in bold to highlight it over the other spectra.
\label{fig_dessart_spec_radius}}
\end{figure}

D14 demonstrated how varying the metallicity affects the model spectra by measuring the equivalent widths (EW) of the various lines as a function of time. 
In their Fig. 6, they compare the EW of the \ion{Fe}{ii}~$\lambda$5018 line with the measured pseudo-EWs (pEW) of a sample of SNe~IIP. 
This reveals a distinct lack of observed SNe~IIP with very metal-poor progenitors, with no SNe in their sample consistent with having a metallicity of less than $0.4\,Z_{\odot}$. 
We measured the pEWs of the \ion{Fe}{ii}~$\lambda$5018 line in the spectra of LSQ13fn, along with SN~1999em, SN~2004et, and SN~2009bw for comparison. 
We also measured the pEWs of the model spectra in an identical manner as in D14. 
These measurements are shown in Fig. \ref{fig_pEW} as a function of epoch, and it is again immediately clear that LSQ13fn is most consistent with the most metal-poor model (m15z2m3).
We note that for SN~2004et and SN~2009bw, the environmental oxygen abundances inferred from relatively local \ion{H}{ii} regions are broadly consistent with the results shown in Fig. \ref{fig_pEW}; for SN~1999em, the \ion{H}{ii} region metallicity is lower than implied by Fig. \ref{fig_pEW}, but neither should be taken literally; we remind the reader that although the synthetic pEWs have been measured from models with a specific set of input parameters, the general agreement with the three SNe is interesting, and future studies incorporating larger numbers of SNe should reveal just how tight the correlation between environmental and SN diagnostics of metallicity really is \citep[e.g.][]{Anderson_2015}.

In addition to a range of initial metallicities, D13 also presented a grid of models that have approximately the same initial conditions with the exception of the radius.
In Fig. \ref{fig_dessart_spec_radius} we compare the 58.3\,d spectrum of LSQ13fn with three models of varied initial radius. 
The flux scale has been matched at the wavelength of H$\alpha$, in the same way as in Fig. \ref{fig_plateau_spec}b. 
The continuum of the model spectrum with the largest radius (m15mlt1, R = 1107 R$_{\odot}$) matches best with LSQ13fn, in that it appears to have less suppression of flux in the blue part of the spectrum. There is a trend of increasing suppression in the blue with decreasing radius. 

D13 discuss the effects of varying progenitor radius on the light curves and spectra of SNe~IIP: models with larger radii produce SNe with brighter plateau luminosities and hotter spectra during the recombination phase due to reduced cooling from expansion. 
These properties appear to be in agreement with those of LSQ13fn; the plateau is relatively luminous, and the temperature is higher (i.e. more blue) compared to typical SNe~IIP (Fig. \ref{fig_plateau_spec}b). 
D14 also note that both the increase of the strength of metal lines and the onset of line blanketing occurs later for models with larger radii, and we have already shown that the evolution of LSQ13fn appears to be somewhat slower than the m15z2m3 model ($R=524\,R_{\odot}$).

The comparison of the spectra with synthetic spectra seems to imply that LSQ13fn lies in a parameter space simultaneously favouring both a low metallicity and a large radius.
In combination these may plausibly explain the weakness of the \ion{Fe}{ii} lines, the lack of lines in the range $\sim$5200-6500\,{\AA} and suppression of flux in the blue by metal line blanketing, and the bright plateau luminosity.
However, as discussed in D14, a red supergiant progenitor at low metallicity will result in a smaller radius, all else being equal.
A more compact star should lead to a fainter plateau, which is in disagreement with LSQ13fn.
The influence of variations in metallicity and radius is complex and difficult to disentangle in this case.

\section{Discussion and conclusions}\label{sec:discussion}

Our data show that while LSQ13fn is consistent with being classified as a SN~IIP, there are several remarkable peculiarities that required us to invoke a number of additional effects that are at play.
Natural variations in red supergiant progenitors, and their resulting explosions as SNe~IIP, are primarily the progenitor radius, the mass of the hydrogen envelope (i.e. the ejecta mass), and the kinetic energy of the explosion.
These parameters can be varied in a continuous manner to account for the majority of diversity in luminosity and colour observed in typical SNe~IIP, but are insufficient to fully explain the peculiar features of LSQ13fn such as the discrepancy from the velocity-luminosity correlation (Fig. \ref{fig_vel_mag}).

Based on our earliest spectra we considered the possibility that ejecta-CSM interaction was responsible for exciting the high-ionisation emission features of \ion{He}{ii}, \ion{N}{iii}, and \ion{C}{iii} in the pre-maximum spectrum.
Although the light curve and post-24.4\,d spectra are reminiscent of SNe~IIP and no strong evidence for ongoing significant CSM interaction is seen, residual thermal energy from the early interaction may go some way to explain the blue colours, high luminosity, slow velocities, and weak lines via veiling.
However the lack of metal lines detected in the wavelength range of $\sim$5200-6500\,{\AA}, in particular, show that interaction can only partly explain the peculiarities.

The comparison with off-the-shelf radiative transfer models pointed towards a combination of low metallicity and large radius, but this combination of parameters needs to be explored further.
At low initial metallicity, the opacity and mass-loss are lower, in turn leading to a more compact progenitor and consequently a SN with a fainter plateau luminosity (D13).
However, there are a number of ways that the progenitor radius may be increased beyond what is expected from metallicity alone, such as a high He-core luminosity, rapid initial rotation, higher progenitor mass, or maybe even tidal effects due to binary interaction.
The low metallicity is supported by the pEW measurements of the \ion{Fe}{ii} $\lambda$5018 line (Fig. \ref{fig_pEW}) but one must bear in mind that the weakness of the lines could be partly explained by veiling.
It seems evident that the quasi-simultaneous contribution of interaction, metallicity, and progenitor radius are all at play in defining the observed properties of LSQ13fn. The diversity in SNe~IIP is further highlighted by our analysis of LSQ13fn which forces us into previously unexplored regions of parameter space.

\begin{acknowledgements}

RK acknowledges support from STFC via ST/L000709/1.
JP and RK acknowledge comments from J. Anderson and C. Guti\'{e}rrez on an earlier 
version.

LD acknowledges financial support from ``Agence Nationale de la Recherche" grant ANR-2011-Blanc-SIMI-5-6-007-01.

This work was partly supported by the EU/FP7 programme via ERC grant number 320360 awarded to G. Gilmore. 

AG-Y's team is supported by the EU/FP7 via ERC grant no.
307260, the Quantum Universe I-Core program by the Israeli
Committee for planning and budgeting and the ISF;
by Minerva and ISF grants; by the Weizmann-UK ``making
connections'' program; and by Kimmel and ARCHES
awards.

SJS acknowledges funding from the European Research Council under the European Union’s Seventh Framework Programme (FP7/2007-2013)/ERC Grant agreement no [291222] and STFC grants ST/I001123/1 and ST/L000709/1.

SB is partially supported by the PRIN-INAF 2014 project Transient Universe: unveiling new types of stellar explosions with PESSTO.

Support for LG is provided by the Ministry of Economy, Development, and Tourism's Millennium Science Initiative through grant IC120009, awarded to The Millennium Institute of Astrophysics, MAS. LG acknowledges support by CONICYT through FONDECYT grant 3140566.

MS acknowledges support from the Royal Society and EU/FP7-
ERC grant number 615929, and STFC via ST/L000679/1.

KT was supported by CONICYT through the FONDECYT grant 3150473, and by the Ministry of Economy, Development, and Tourism’s Millennium Science Initiative through grant IC12009, awarded to the Millennium Institute of Astrophysics, MAS.

This work is based (in part) on observations collected at the European Organisation for Astronomical Research in the Southern Hemisphere, Chile as part of PESSTO, (the Public ESO Spectroscopic Survey for Transient Objects) ESO programme 188.D-3003/191.D-0935, and as part of programme 092.D-0586.

This work utilizes data from the 40-inch ESO Schmidt Telescope at the La Silla Observatory in Chile with the large area QUEST camera built at Yale University and Indiana University.

The Liverpool Telescope is operated on the island of La Palma by Liverpool John Moores University in the Spanish Observatorio del Roque de los Muchachos of the Instituto de Astrofisica de Canarias with financial support from the UK Science and Technology Facilities Council (proposal ID: PL12B05, P.I. R. Kotak).

The William Herschel Telescope is operated on the island of La Palma by the Isaac Newton Group in the Spanish Observatorio del Roque de los Muchachos of the Instituto de Astrofísica de Canarias (programme W/2013A/20).

This research has made use of the NASA/IPAC Extragalactic Database (NED) which is operated by the Jet Propulsion Laboratory, California Institute of Technology, under contract with the National Aeronautics and Space Administration. 

\end{acknowledgements}

\bibliographystyle{aa}
\bibliography{bibtex}

\begin{thebibliography}{67}
\expandafter\ifx\csname natexlab\endcsname\relax\def\natexlab#1{#1}\fi

\bibitem[{{Anderson} {et~al.}(2014{\natexlab{a}}){Anderson}, {Dessart},
  {Gutierrez}, {Hamuy}, {Morrell}, {Phillips}, {Folatelli}, {Stritzinger},
  {Freedman}, {Gonz{\'a}lez-Gait{\'a}n}, {McCarthy}, {Suntzeff}, \&
  {Thomas-Osip}}]{Anderson_2014b}
{Anderson}, J.~P., {Dessart}, L., {Gutierrez}, C.~P., {et~al.}
  2014{\natexlab{a}}, \mnras, 441, 671

\bibitem[{{Anderson} {et~al.}(2014{\natexlab{b}}){Anderson},
  {Gonz{\'a}lez-Gait{\'a}n}, {Hamuy}, {Guti{\'e}rrez}, {Stritzinger}, {Olivares
  E.}, {Phillips}, {Schulze}, {Antezana}, {Bolt}, {Campillay}, {Castell{\'o}n},
  {Contreras}, {de Jaeger}, {Folatelli}, {F{\"o}rster}, {Freedman},
  {Gonz{\'a}lez}, {Hsiao}, {Krzemi{\'n}ski}, {Krisciunas}, {Maza}, {McCarthy},
  {Morrell}, {Persson}, {Roth}, {Salgado}, {Suntzeff}, \&
  {Thomas-Osip}}]{Anderson_2014}
{Anderson}, J.~P., {Gonz{\'a}lez-Gait{\'a}n}, S., {Hamuy}, M., {et~al.}
  2014{\natexlab{b}}, \apj, 786, 67

\bibitem[{{Anderson} {et~al.}(2015){Anderson}, {Guti{\'e}rrez}, \&
  {Dessart}}]{Anderson_2015}
{Anderson}, J.~P., {Guti{\'e}rrez}, C.~P., \& {Dessart}, L. 2015, arXiv,
  arXiv:1510.04271

\bibitem[{{Arcavi} {et~al.}(2012){Arcavi}, {Gal-Yam}, {Cenko}, {Fox},
  {Leonard}, {Moon}, {Sand}, {Soderberg}, {Kiewe}, {Yaron}, {Becker}, {Scheps},
  {Birenbaum}, {Chamudot}, \& {Zhou}}]{Arcavi_2012}
{Arcavi}, I., {Gal-Yam}, A., {Cenko}, S.~B., {et~al.} 2012, \apjl, 756, L30

\bibitem[{{Arnett} {et~al.}(1989){Arnett}, {Bahcall}, {Kirshner}, \&
  {Woosley}}]{Arnett_1989b}
{Arnett}, W.~D., {Bahcall}, J.~N., {Kirshner}, R.~P., \& {Woosley}, S.~E. 1989,
  \araa, 27, 629

\bibitem[{{Arnett} \& {Fu}(1989)}]{Arnett_1989}
{Arnett}, W.~D. \& {Fu}, A. 1989, \apj, 340, 396

\bibitem[{{Asplund} {et~al.}(2009){Asplund}, {Grevesse}, {Sauval}, \&
  {Scott}}]{Asplund_2009}
{Asplund}, M., {Grevesse}, N., {Sauval}, A.~J., \& {Scott}, P. 2009, \araa, 47,
  481

\bibitem[{{Baltay} {et~al.}(2013){Baltay}, {Rabinowitz}, {Hadjiyska}, {Walker},
  {Nugent}, {Coppi}, {Ellman}, {Feindt}, {McKinnon}, {Horowitz}, \&
  {Effron}}]{Baltay_2013}
{Baltay}, C., {Rabinowitz}, D., {Hadjiyska}, E., {et~al.} 2013, \pasp, 125, 683

\bibitem[{{Barbon} {et~al.}(1979){Barbon}, {Ciatti}, \& {Rosino}}]{Barbon_1979}
{Barbon}, R., {Ciatti}, F., \& {Rosino}, L. 1979, \aap, 72, 287

\bibitem[{{Baron} {et~al.}(2000){Baron}, {Branch}, {Hauschildt}, {Filippenko},
  {Kirshner}, {Challis}, {Jha}, {Chevalier}, {Fransson}, {Lundqvist},
  {Garnavich}, {Leibundgut}, {McCray}, {Michael}, {Panagia}, {Phillips}, {Pun},
  {Schmidt}, {Sonneborn}, {Suntzeff}, {Wang}, \& {Wheeler}}]{Baron_2000}
{Baron}, E., {Branch}, D., {Hauschildt}, P.~H., {et~al.} 2000, \apj, 545, 444

\bibitem[{{Bersten} {et~al.}(2011){Bersten}, {Benvenuto}, \&
  {Hamuy}}]{Bersten_2011}
{Bersten}, M.~C., {Benvenuto}, O., \& {Hamuy}, M. 2011, \apj, 729, 61

\bibitem[{{Bouchet} {et~al.}(1991){Bouchet}, {Danziger}, \&
  {Lucy}}]{Bouchet_1991}
{Bouchet}, P., {Danziger}, I.~J., \& {Lucy}, L.~B. 1991, \aj, 102, 1135

\bibitem[{{Dessart} {et~al.}(2014){Dessart}, {Gutierrez}, {Hamuy}, {Hillier},
  {Lanz}, {Anderson}, {Folatelli}, {Freedman}, {Ley}, {Morrell}, {Persson},
  {Phillips}, {Stritzinger}, \& {Suntzeff}}]{Dessart_2014}
{Dessart}, L., {Gutierrez}, C.~P., {Hamuy}, M., {et~al.} 2014, \mnras, 440,
  1856

\bibitem[{{Dessart} {et~al.}(2013){Dessart}, {Hillier}, {Waldman}, \&
  {Livne}}]{Dessart_2013}
{Dessart}, L., {Hillier}, D.~J., {Waldman}, R., \& {Livne}, E. 2013, \mnras,
  433, 1745

\bibitem[{{Dessart} {et~al.}(2010){Dessart}, {Livne}, \&
  {Waldman}}]{Dessart_2010}
{Dessart}, L., {Livne}, E., \& {Waldman}, R. 2010, \mnras, 408, 827

\bibitem[{{Elmhamdi} {et~al.}(2003){Elmhamdi}, {Danziger}, {Chugai},
  {Pastorello}, {Turatto}, {Cappellaro}, {Altavilla}, {Benetti}, {Patat}, \&
  {Salvo}}]{elmhamdi_2003}
{Elmhamdi}, A., {Danziger}, I.~J., {Chugai}, N., {et~al.} 2003, \mnras, 338,
  939

\bibitem[{{Faran} {et~al.}(2014){Faran}, {Poznanski}, {Filippenko}, {Chornock},
  {Foley}, {Ganeshalingam}, {Leonard}, {Li}, {Modjaz}, {Serduke}, \&
  {Silverman}}]{Faran_2014}
{Faran}, T., {Poznanski}, D., {Filippenko}, A.~V., {et~al.} 2014, \mnras, 445,
  554

\bibitem[{{Fassia} {et~al.}(2001){Fassia}, {Meikle}, {Chugai}, {Geballe},
  {Lundqvist}, {Walton}, {Pollacco}, {Veilleux}, {Wright}, {Pettini}, {Kerr},
  {Puchnarewicz}, {Puxley}, {Irwin}, {Packham}, {Smartt}, \&
  {Harmer}}]{fassia_2001}
{Fassia}, A., {Meikle}, W.~P.~S., {Chugai}, N., {et~al.} 2001, \mnras, 325, 907

\bibitem[{{Filippenko}(1997)}]{filippenko_1997}
{Filippenko}, A.~V. 1997, \araa, 35, 309

\bibitem[{{Fraser} {et~al.}(2015){Fraser}, {Kotak}, {Pastorello}, {Jerkstrand},
  {Smartt}, {Chen}, {Childress}, {Gilmore}, {Inserra}, {Kankare}, {Margheim},
  {Mattila}, {Valenti}, {Ashall}, {Benetti}, {Botticella}, {Bauer}, {Campbell},
  {Elias-Rosa}, {Fleury}, {Gal-Yam}, {Hachinger}, {Howell}, {Le Guillou},
  {L{\'e}get}, {Morales-Garoffolo}, {Polshaw}, {Spiro}, {Sullivan},
  {Taubenberger}, {Turatto}, {Walker}, {Young}, \& {Zhang}}]{Fraser_2015}
{Fraser}, M., {Kotak}, R., {Pastorello}, A., {et~al.} 2015, \mnras, 453, 3886

\bibitem[{{Gal-Yam} {et~al.}(2014){Gal-Yam}, {Arcavi}, {Ofek}, {Ben-Ami},
  {Cenko}, {Kasliwal}, {Cao}, {Yaron}, {Tal}, {Silverman}, {Horesh}, {De Cia},
  {Taddia}, {Sollerman}, {Perley}, {Vreeswijk}, {Kulkarni}, {Nugent},
  {Filippenko}, \& {Wheeler}}]{Gal-Yam_2014}
{Gal-Yam}, A., {Arcavi}, I., {Ofek}, E.~O., {et~al.} 2014, \nat, 509, 471

\bibitem[{{Gal-Yam} {et~al.}(2013){Gal-Yam}, {Mazzali}, {Manulis}, \&
  {Bishop}}]{Gal-Yam_2013}
{Gal-Yam}, A., {Mazzali}, P.~A., {Manulis}, I., \& {Bishop}, D. 2013, \pasp,
  125, 749

\bibitem[{{Gall} {et~al.}(2015){Gall}, {Polshaw}, {Kotak}, {Jerkstrand},
  {Leibundgut}, {Rabinowitz}, {Sollerman}, {Sullivan}, {Smartt}, {Anderson},
  {Benetti}, {Baltay}, {Feindt}, {Fraser}, {Gonz{\'a}lez-Gait{\'a}n},
  {Inserra}, {Maguire}, {McKinnon}, {Valenti}, \& {Young}}]{Gall_2015}
{Gall}, E.~E.~E., {Polshaw}, J., {Kotak}, R., {et~al.} 2015, \aap, 582, A3

\bibitem[{{Gonz{\'a}lez-Gait{\'a}n} {et~al.}(2015){Gonz{\'a}lez-Gait{\'a}n},
  {Tominaga}, {Molina}, {Galbany}, {Bufano}, {Anderson}, {Gutierrez},
  {F{\"o}rster}, {Pignata}, {Bersten}, {Howell}, {Sullivan}, {Carlberg}, {de
  Jaeger}, {Hamuy}, {Baklanov}, \& {Blinnikov}}]{Gonzalez_2015}
{Gonz{\'a}lez-Gait{\'a}n}, S., {Tominaga}, N., {Molina}, J., {et~al.} 2015,
  \mnras, 451, 2212

\bibitem[{{Gr{\"a}fener} \& {Vink}(2015)}]{Grafener_2015}
{Gr{\"a}fener}, G. \& {Vink}, J.~S. 2015, arXiv, arXiv:1510.00013

\bibitem[{{Groh}(2014)}]{Groh_2014}
{Groh}, J.~H. 2014, \aap, 572, L11

\bibitem[{{Hamuy}(2003)}]{Hamuy_2003}
{Hamuy}, M. 2003, \apj, 582, 905

\bibitem[{{Hamuy} \& {Pinto}(2002)}]{Hamuy_2002}
{Hamuy}, M. \& {Pinto}, P.~A. 2002, \apjl, 566, L63

\bibitem[{{Hamuy} {et~al.}(2001){Hamuy}, {Pinto}, {Maza}, {Suntzeff},
  {Phillips}, {Eastman}, {Smith}, {Corbally}, {Burstein}, {Li}, {Ivanov},
  {Moro-Martin}, {Strolger}, {de Souza}, {dos Anjos}, {Green}, {Pickering},
  {Gonz{\'a}lez}, {Antezana}, {Wischnjewsky}, {Galaz}, {Roth}, {Persson}, \&
  {Schommer}}]{hamuy_2001}
{Hamuy}, M., {Pinto}, P.~A., {Maza}, J., {et~al.} 2001, \apj, 558, 615

\bibitem[{{Henry} \& {Worthey}(1999)}]{Henry_1999}
{Henry}, R.~B.~C. \& {Worthey}, G. 1999, \pasp, 111, 919

\bibitem[{{Inserra} {et~al.}(2013){Inserra}, {Pastorello}, {Turatto}, {Pumo},
  {Benetti}, {Cappellaro}, {Botticella}, {Bufano}, {Elias-Rosa}, {Harutyunyan},
  {Taubenberger}, {Valenti}, \& {Zampieri}}]{Inserra_2013}
{Inserra}, C., {Pastorello}, A., {Turatto}, M., {et~al.} 2013, \aap, 555, A142

\bibitem[{{Inserra} {et~al.}(2012){Inserra}, {Turatto}, {Pastorello}, {Pumo},
  {Baron}, {Benetti}, {Cappellaro}, {Taubenberger}, {Bufano}, {Elias-Rosa},
  {Zampieri}, {Harutyunyan}, {Moskvitin}, {Nissinen}, {Stanishev}, {Tsvetkov},
  {Hentunen}, {Komarova}, {Pavlyuk}, {Sokolov}, \& {Sokolova}}]{Inserra_2012}
{Inserra}, C., {Turatto}, M., {Pastorello}, A., {et~al.} 2012, \mnras, 422,
  1122

\bibitem[{{Kasen} \& {Woosley}(2009)}]{Kasen_2009}
{Kasen}, D. \& {Woosley}, S.~E. 2009, \apj, 703, 2205

\bibitem[{{Leonard} {et~al.}(2003){Leonard}, {Kanbur}, {Ngeow}, \&
  {Tanvir}}]{leonard_2003}
{Leonard}, D.~C., {Kanbur}, S.~M., {Ngeow}, C.~C., \& {Tanvir}, N.~R. 2003,
  \apj, 594, 247

\bibitem[{{Li} {et~al.}(2011){Li}, {Leaman}, {Chornock}, {Filippenko},
  {Poznanski}, {Ganeshalingam}, {Wang}, {Modjaz}, {Jha}, {Foley}, \&
  {Smith}}]{Li_2011}
{Li}, W., {Leaman}, J., {Chornock}, R., {et~al.} 2011, \mnras, 412, 1441

\bibitem[{{Liu} {et~al.}(2000){Liu}, {Hu}, {Hang}, {Qiu}, {Zhu}, \&
  {Qiao}}]{liu_2000}
{Liu}, Q.-Z., {Hu}, J.-Y., {Hang}, H.-R., {et~al.} 2000, \aaps, 144, 219

\bibitem[{{Maguire} {et~al.}(2010){Maguire}, {Di Carlo}, {Smartt},
  {Pastorello}, {Tsvetkov}, {Benetti}, {Spiro}, {Arkharov}, {Beccari},
  {Botticella}, {Cappellaro}, {Cristallo}, {Dolci}, {Elias-Rosa}, {Fiaschi},
  {Gorshanov}, {Harutyunyan}, {Larionov}, {Navasardyan}, {Pietrinferni},
  {Raimondo}, {di Rico}, {Valenti}, {Valentini}, \& {Zampieri}}]{maguire_2010}
{Maguire}, K., {Di Carlo}, E., {Smartt}, S.~J., {et~al.} 2010, \mnras, 404, 981

\bibitem[{{Minkowski}(1941)}]{Minkowski_1941}
{Minkowski}, R. 1941, \pasp, 53, 224

\bibitem[{{Modjaz} {et~al.}(2011){Modjaz}, {Kewley}, {Bloom}, {Filippenko},
  {Perley}, \& {Silverman}}]{Modjaz_2011}
{Modjaz}, M., {Kewley}, L., {Bloom}, J.~S., {et~al.} 2011, \apjl, 731, L4

\bibitem[{{Nugent} {et~al.}(2006){Nugent}, {Sullivan}, {Ellis}, {Gal-Yam},
  {Leonard}, {Howell}, {Astier}, {Carlberg}, {Conley}, {Fabbro}, {Fouchez},
  {Neill}, {Pain}, {Perrett}, {Pritchet}, \& {Regnault}}]{Nugent_2006}
{Nugent}, P., {Sullivan}, M., {Ellis}, R., {et~al.} 2006, \apj, 645, 841

\bibitem[{{Nugent} {et~al.}(2011){Nugent}, {Sullivan}, {Cenko}, {Thomas},
  {Kasen}, {Howell}, {Bersier}, {Bloom}, {Kulkarni}, {Kandrashoff},
  {Filippenko}, {Silverman}, {Marcy}, {Howard}, {Isaacson}, {Maguire},
  {Suzuki}, {Tarlton}, {Pan}, {Bildsten}, {Fulton}, {Parrent}, {Sand},
  {Podsiadlowski}, {Bianco}, {Dilday}, {Graham}, {Lyman}, {James}, {Kasliwal},
  {Law}, {Quimby}, {Hook}, {Walker}, {Mazzali}, {Pian}, {Ofek}, {Gal-Yam}, \&
  {Poznanski}}]{Nugent_2011}
{Nugent}, P.~E., {Sullivan}, M., {Cenko}, S.~B., {et~al.} 2011, \nat, 480, 344

\bibitem[{{Oke} \& {Searle}(1974)}]{Oke_1974}
{Oke}, J.~B. \& {Searle}, L. 1974, \araa, 12, 315

\bibitem[{{Pastorello} {et~al.}(2015){Pastorello}, {Benetti}, {Brown},
  {Tsvetkov}, {Inserra}, {Taubenberger}, {Tomasella}, {Fraser}, {Rich},
  {Botticella}, {Bufano}, {Cappellaro}, {Ergon}, {Gorbovskoy}, {Harutyunyan},
  {Huang}, {Kotak}, {Lipunov}, {Magill}, {Miluzio}, {Morrell}, {Ochner},
  {Smartt}, {Sollerman}, {Spiro}, {Stritzinger}, {Turatto}, {Valenti}, {Wang},
  {Wright}, {Yurkov}, {Zampieri}, \& {Zhang}}]{Pastorello_2015}
{Pastorello}, A., {Benetti}, S., {Brown}, P.~J., {et~al.} 2015, \mnras, 449,
  1921

\bibitem[{{Pastorello} {et~al.}(2009){Pastorello}, {Valenti}, {Zampieri},
  {Navasardyan}, {Taubenberger}, {Smartt}, {Arkharov}, {B{\"a}rnbantner},
  {Barwig}, {Benetti}, {Birtwhistle}, {Botticella}, {Cappellaro}, {Del
  Principe}, {di Mille}, {di Rico}, {Dolci}, {Elias-Rosa}, {Efimova},
  {Fiedler}, {Harutyunyan}, {H{\"o}flich}, {Kloehr}, {Larionov}, {Lorenzi},
  {Maund}, {Napoleone}, {Ragni}, {Richmond}, {Ries}, {Spiro}, {Temporin},
  {Turatto}, \& {Wheeler}}]{Pastorello_2009}
{Pastorello}, A., {Valenti}, S., {Zampieri}, L., {et~al.} 2009, \mnras, 394,
  2266

\bibitem[{{Pastorello} {et~al.}(2004){Pastorello}, {Zampieri}, {Turatto},
  {Cappellaro}, {Meikle}, {Benetti}, {Branch}, {Baron}, {Patat}, {Armstrong},
  {Altavilla}, {Salvo}, \& {Riello}}]{Pastorello_2004}
{Pastorello}, A., {Zampieri}, L., {Turatto}, M., {et~al.} 2004, \mnras, 347, 74

\bibitem[{{Patat} {et~al.}(1994){Patat}, {Barbon}, {Cappellaro}, \&
  {Turatto}}]{Patat_1994}
{Patat}, F., {Barbon}, R., {Cappellaro}, E., \& {Turatto}, M. 1994, \aap, 282,
  731

\bibitem[{{Pettini} \& {Pagel}(2004)}]{pettini_pagel_2004}
{Pettini}, M. \& {Pagel}, B.~E.~J. 2004, \mnras, 348, L59

\bibitem[{{Polshaw} {et~al.}(2015){Polshaw}, {Kotak}, {Chambers}, {Smartt},
  {Taubenberger}, {Kromer}, {Gall}, {Hillebrandt}, {Huber}, {Smith}, \&
  {Wainscoat}}]{Polshaw_2015}
{Polshaw}, J., {Kotak}, R., {Chambers}, K.~C., {et~al.} 2015, \aap, 580, L15

\bibitem[{{Poznanski} {et~al.}(2009){Poznanski}, {Butler}, {Filippenko},
  {Ganeshalingam}, {Li}, {Bloom}, {Chornock}, {Foley}, {Nugent}, {Silverman},
  {Cenko}, {Gates}, {Leonard}, {Miller}, {Modjaz}, {Serduke}, {Smith}, {Swift},
  \& {Wong}}]{Poznanski_2009}
{Poznanski}, D., {Butler}, N., {Filippenko}, A.~V., {et~al.} 2009, \apj, 694,
  1067

\bibitem[{{Poznanski} {et~al.}(2011){Poznanski}, {Ganeshalingam}, {Silverman},
  \& {Filippenko}}]{Poznanski_2011}
{Poznanski}, D., {Ganeshalingam}, M., {Silverman}, J.~M., \& {Filippenko},
  A.~V. 2011, \mnras, 415, L81

\bibitem[{{Poznanski} {et~al.}(2012){Poznanski}, {Prochaska}, \&
  {Bloom}}]{Poznanski_2012}
{Poznanski}, D., {Prochaska}, J.~X., \& {Bloom}, J.~S. 2012, \mnras, 426, 1465

\bibitem[{{Pozzo} {et~al.}(2004){Pozzo}, {Meikle}, {Fassia}, {Geballe},
  {Lundqvist}, {Chugai}, \& {Sollerman}}]{Pozzo_2004}
{Pozzo}, M., {Meikle}, W.~P.~S., {Fassia}, A., {et~al.} 2004, \mnras, 352, 457

\bibitem[{{Pritchard} {et~al.}(2012){Pritchard}, {Roming}, {Brown}, {Kuin},
  {Bayless}, {Holland}, {Immler}, {Milne}, \& {Oates}}]{Pritchard_2012}
{Pritchard}, T.~A., {Roming}, P.~W.~A., {Brown}, P.~J., {et~al.} 2012, \apj,
  750, 128

\bibitem[{{Sahu} {et~al.}(2006){Sahu}, {Anupama}, {Srividya}, \&
  {Muneer}}]{sahu_2006}
{Sahu}, D.~K., {Anupama}, G.~C., {Srividya}, S., \& {Muneer}, S. 2006, \mnras,
  372, 1315

\bibitem[{{S{\'a}nchez} {et~al.}(2014){S{\'a}nchez}, {Rosales-Ortega},
  {Iglesias-P{\'a}ramo}, {Moll{\'a}}, {Barrera-Ballesteros}, {Marino},
  {P{\'e}rez}, {S{\'a}nchez-Blazquez}, {Gonz{\'a}lez Delgado}, {Cid Fernandes},
  {de Lorenzo-C{\'a}ceres}, {Mendez-Abreu}, {Galbany}, {Falcon-Barroso},
  {Miralles-Caballero}, {Husemann}, {Garc{\'{\i}}a-Benito}, {Mast}, {Walcher},
  {Gil de Paz}, {Garc{\'{\i}}a-Lorenzo}, {Jungwiert}, {V{\'{\i}}lchez},
  {J{\'{\i}}lkov{\'a}}, {Lyubenova}, {Cortijo-Ferrero}, {D{\'{\i}}az},
  {Wisotzki}, {M{\'a}rquez}, {Bland-Hawthorn}, {Ellis}, {van de Ven}, {Jahnke},
  {Papaderos}, {Gomes}, {Mendoza}, \& {L{\'o}pez-S{\'a}nchez}}]{Sanchez_2014}
{S{\'a}nchez}, S.~F., {Rosales-Ortega}, F.~F., {Iglesias-P{\'a}ramo}, J.,
  {et~al.} 2014, \aap, 563, A49

\bibitem[{{Sanders} {et~al.}(2015){Sanders}, {Soderberg}, {Gezari},
  {Betancourt}, {Chornock}, {Berger}, {Foley}, {Challis}, {Drout}, {Kirshner},
  {Lunnan}, {Marion}, {Margutti}, {McKinnon}, {Milisavljevic}, {Narayan},
  {Rest}, {Kankare}, {Mattila}, {Smartt}, {Huber}, {Burgett}, {Draper},
  {Hodapp}, {Kaiser}, {Kudritzki}, {Magnier}, {Metcalfe}, {Morgan}, {Price},
  {Tonry}, {Wainscoat}, \& {Waters}}]{Sanders_2015}
{Sanders}, N.~E., {Soderberg}, A.~M., {Gezari}, S., {et~al.} 2015, \apj, 799,
  208

\bibitem[{{Schlafly} \& {Finkbeiner}(2011)}]{Schlafly_2011}
{Schlafly}, E.~F. \& {Finkbeiner}, D.~P. 2011, \apj, 737, 103

\bibitem[{{Schlegel}(1990)}]{Schlegel_1990}
{Schlegel}, E.~M. 1990, \mnras, 244, 269

\bibitem[{{Shivvers} {et~al.}(2015){Shivvers}, {Groh}, {Mauerhan}, {Fox},
  {Leonard}, \& {Filippenko}}]{Shivvers_2014}
{Shivvers}, I., {Groh}, J.~H., {Mauerhan}, J.~C., {et~al.} 2015, \apj, 806, 213

\bibitem[{{Skrutskie} {et~al.}(2006){Skrutskie}, {Cutri}, {Stiening},
  {Weinberg}, {Schneider}, {Carpenter}, {Beichman}, {Capps}, {Chester},
  {Elias}, {Huchra}, {Liebert}, {Lonsdale}, {Monet}, {Price}, {Seitzer},
  {Jarrett}, {Kirkpatrick}, {Gizis}, {Howard}, {Evans}, {Fowler}, {Fullmer},
  {Hurt}, {Light}, {Kopan}, {Marsh}, {McCallon}, {Tam}, {Van Dyk}, \&
  {Wheelock}}]{Skrutskie_2006}
{Skrutskie}, M.~F., {Cutri}, R.~M., {Stiening}, R., {et~al.} 2006, \aj, 131,
  1163

\bibitem[{{Smartt} {et~al.}(2009){Smartt}, {Eldridge}, {Crockett}, \&
  {Maund}}]{Smartt_2009}
{Smartt}, S.~J., {Eldridge}, J.~J., {Crockett}, R.~M., \& {Maund}, J.~R. 2009,
  \mnras, 395, 1409

\bibitem[{{Smartt} {et~al.}(2015){Smartt}, {Valenti}, {Fraser}, {Inserra},
  {Young}, {Sullivan}, {Pastorello}, {Benetti}, {Gal-Yam}, {Knapic},
  {Molinaro}, {Smareglia}, {Smith}, {Taubenberger}, {Yaron}, {Anderson},
  {Ashall}, {Balland}, {Baltay}, {Barbarino}, {Bauer}, {Baumont}, {Bersier},
  {Blagorodnova}, {Bongard}, {Botticella}, {Bufano}, {Bulla}, {Cappellaro},
  {Campbell}, {Cellier-Holzem}, {Chen}, {Childress}, {Clocchiatti},
  {Contreras}, {Dall'Ora}, {Danziger}, {de Jaeger}, {De Cia}, {Della Valle},
  {Dennefeld}, {Elias-Rosa}, {Elman}, {Feindt}, {Fleury}, {Gall},
  {Gonzalez-Gaitan}, {Galbany}, {Morales Garoffolo}, {Greggio}, {Guillou},
  {Hachinger}, {Hadjiyska}, {Hage}, {Hillebrandt}, {Hodgkin}, {Hsiao}, {James},
  {Jerkstrand}, {Kangas}, {Kankare}, {Kotak}, {Kromer}, {Kuncarayakti},
  {Leloudas}, {Lundqvist}, {Lyman}, {Hook}, {Maguire}, {Manulis}, {Margheim},
  {Mattila}, {Maund}, {Mazzali}, {McCrum}, {McKinnon}, {Moreno-Raya},
  {Nicholl}, {Nugent}, {Pain}, {Pignata}, {Phillips}, {Polshaw}, {Pumo},
  {Rabinowitz}, {Reilly}, {Romero-Ca{\~n}izales}, {Scalzo}, {Schmidt},
  {Schulze}, {Sim}, {Sollerman}, {Taddia}, {Tartaglia}, {Terreran},
  {Tomasella}, {Turatto}, {Walker}, {Walton}, {Wyrzykowski}, {Yuan}, \&
  {Zampieri}}]{Smartt_2015}
{Smartt}, S.~J., {Valenti}, S., {Fraser}, M., {et~al.} 2015, \aap, 579, A40

\bibitem[{{Sollerman} {et~al.}(2013){Sollerman}, {Taddia}, {Ergon}, {Leloudas},
  {Elias-Rosa}, {Scalzo}, {Baltay}, {Ellman}, {Hadjiyska}, {McKinnon},
  {Rabinowitz}, {Walker}, {Feindt}, {Kowalski}, {Nugent}, {Benetti},
  {Pastorello}, {Valenti}, {Taubenberger}, {Smartt}, {Smith}, {Young},
  {Sullivan}, {Gal-Yam}, \& {Yaron}}]{Sollerman_2013}
{Sollerman}, J., {Taddia}, F., {Ergon}, M., {et~al.} 2013, The Astronomer's
  Telegram, 4731, 1

\bibitem[{{Spiro} {et~al.}(2014){Spiro}, {Pastorello}, {Pumo}, {Zampieri},
  {Turatto}, {Smartt}, {Benetti}, {Cappellaro}, {Valenti}, {Agnoletto},
  {Altavilla}, {Aoki}, {Brocato}, {Corsini}, {Di Cianno}, {Elias-Rosa},
  {Hamuy}, {Enya}, {Fiaschi}, {Folatelli}, {Desidera}, {Harutyunyan}, {Howell},
  {Kawka}, {Kobayashi}, {Leibundgut}, {Minezaki}, {Navasardyan}, {Nomoto},
  {Mattila}, {Pietrinferni}, {Pignata}, {Raimondo}, {Salvo}, {Schmidt},
  {Sollerman}, {Spyromilio}, {Taubenberger}, {Valentini}, {Vennes}, \&
  {Yoshii}}]{Spiro_2014}
{Spiro}, S., {Pastorello}, A., {Pumo}, M.~L., {et~al.} 2014, \mnras, 439, 2873

\bibitem[{{Turatto} {et~al.}(2003){Turatto}, {Benetti}, \&
  {Cappellaro}}]{Turatto_2003}
{Turatto}, M., {Benetti}, S., \& {Cappellaro}, E. 2003, in From Twilight to
  Highlight: The Physics of Supernovae, ed. W.~{Hillebrandt} \&
  B.~{Leibundgut}, 200

\bibitem[{{van Dokkum}(2001)}]{dokkum_2001}
{van Dokkum}, P.~G. 2001, \pasp, 113, 1420

\bibitem[{{Yaron} \& {Gal-Yam}(2012)}]{Yaron_2012}
{Yaron}, O. \& {Gal-Yam}, A. 2012, \pasp, 124, 668

\end{thebibliography}

\appendix

\section{Tables}

\begin{table*}
\setlength{\tabcolsep}{3pt}
\caption{Optical $UBVRI$ photometry of LSQ13fn. Uncertainties are given in parentheses. The magnitudes have not been corrected for reddening.}
\label{table:photometry_ubvri} 
 \centering
 \begin{tabular}{lcccccccc}
 \hline\hline 
Date	&	MJD	&	Epoch*	&	$U$	&	$B$	&	$V$	&	$R$	&	$I$	&	Telescope	\\
yyyy mm dd	&	&	&	&	&	&	&	&	\\
 \hline
2012 12 23	&	56284.3	&	$-20.6$	&	\ldots	&	\ldots	&	$>$21.95	&	\ldots	&	\ldots	&	LSQ	\\
2013 01 10	&	56302.2	&	\phs$-3.8$	&	\ldots	&	\ldots	&	20.35 (0.22)	&	\ldots	&	\ldots	&	LSQ	\\
2013 01 12	&	56304.2	&	\phs$-1.9$	&	\ldots	&	\ldots	&	19.65 (0.20)	&	\ldots	&	\ldots	&	LSQ	\\
2013 01 13	&	56305.3	&	\phs$-0.8$	&	18.59 (0.06)	&	19.63 (0.05)	&	19.75 (0.06)	&	19.72 (0.06)	&	\ldots	&	NTT	\\
2013 01 14	&	56306.2	&	\phs\phs0.0	&	\ldots	&	\ldots	&	19.48 (0.18)	&	\ldots	&	\ldots	&	LSQ	\\
2013 01 16	&	56308.3	&	\phs\phs1.9	&	\ldots	&	\ldots	&	19.55 (0.22)	&	\ldots	&	\ldots	&	LSQ	\\
2013 01 20	&	56312.3	&	\phs\phs5.7	&	18.63 (0.11)	&	19.64 (0.06)	&	19.55 (0.05)	&	19.47 (0.05)	&	\ldots	&	NTT	\\
2013 01 22	&	56314.3	&	\phs\phs7.6	&	18.90 (0.08)	&	19.62 (0.07)	&	19.63 (0.07)	&	19.61 (0.07)	&	\ldots&	NTT	\\
2013 01 30	&	56322.3	&	\phs15.1	&	19.23 (0.08)	&	19.91 (0.05)	&	19.74 (0.06)	&	19.62 (0.06)	&	\ldots	&	NTT	\\
2013 02 09	&	56332.2	&	\phs24.4	&	\ldots	&	\ldots	&	19.70 (0.08)	&	\ldots	&	\ldots	&	NTT	\\
2013 02 11	&	56334.2	&	\phs26.3	&	\ldots	&	20.22 (0.05)	&	19.91 (0.05)	&	19.78 (0.05)	&	19.59 (0.08)	&	SMARTS	\\
2013 02 14	&	56337.2	&	\phs29.2	&	\ldots	&	20.26 (0.05)	&	20.00 (0.04)	&	19.75 (0.04)	&	19.55 (0.20)	&	SMARTS	\\
2013 02 17	&	56340.3	&	\phs32.0	&	\ldots	&	20.35 (0.05)	&	20.00 (0.04)	&	19.74 (0.05)	&	19.57 (0.06)	&	SMARTS	\\
2013 02 21	&	56344.2	&	\phs35.7	&	\ldots	&	20.36 (0.07)	&	19.98 (0.09)	&	19.69 (0.07)	&	19.44 (0.11)	&	SMARTS	\\
2013 03 03	&	56354.1	&	\phs45.0	&	\ldots	&	\ldots	&	\ldots	&	19.55 (0.10)	&	\ldots	&	NTT	\\
2013 03 06	&	56357.2	&	\phs48.0	&	\ldots	&	20.45 (0.17)	&	20.13 (0.17)	&	19.82 (0.10)	&	\ldots	&	SMARTS	\\
2013 03 09	&	56360.2	&	\phs50.8	&	\ldots	&	20.73 (0.05)	&	20.06 (0.08)	&	19.78 (0.09)	&	19.73 (0.19)	&	SMARTS	\\
2013 03 12	&	56363.2	&	\phs53.7	&	\ldots	&	20.76 (0.07)	&	20.19 (0.06)	&	19.85 (0.09)	&	19.65 (0.10)	&	SMARTS	\\
2013 03 13	&	56364.3	&	\phs54.6	&	\ldots	&	\ldots	&	20.03 (0.10)	&	\ldots	&	\ldots	&	NTT	\\
2013 03 16	&	56367.3	&	\phs57.4	&	\ldots	&	21.07 (0.10)	&	20.16 (0.09)	&	19.77 (0.05)	&	19.51 (0.12)	&	SMARTS	\\
2013 03 17	&	56368.2	&	\phs58.3	&	\ldots	&	\ldots	&	20.18 (0.06)	&	\ldots	&	\ldots	&	NTT	\\
2013 03 20	&	56371.3	&	\phs61.2	&	\ldots	&	21.05 (0.09)	&	20.19 (0.05)	&	19.92 (0.05)	&	19.58 (0.10)	&	SMARTS	\\
2013 03 24	&	56375.2	&	\phs64.9	&	\ldots	&	21.34 (0.14)	&	20.21 (0.14)	&	19.92 (0.09)	&	19.70 (0.10)	&	SMARTS	\\
2013 03 31	&	56382.0	&	\phs71.3	&	\ldots	&	21.31 (0.23)	&	20.30 (0.27)	&	\ldots	&	\ldots	&	LT	\\
2013 04 02	&	56384.1	&	\phs73.3	&	\ldots	&	\ldots	&	20.25 (0.10)	&	19.94 (0.09)	&	\ldots	&	SMARTS	\\
2013 04 05	&	56387.2	&	\phs76.2	&	\ldots	&	\ldots	&	20.31 (0.08)	&	\ldots	&	\ldots	&	NTT	\\
2013 04 14	&	56396.1	&	\phs84.5	&	\ldots	&	21.63 (0.11)	&	20.54 (0.08)	&	20.17 (0.05)	&	\ldots	&	NTT	\\   
2013 04 15	&	56397.0	&	\phs85.4	&	\ldots	&	21.94 (0.24)	&	20.48 (0.12)	&	\ldots	&	\ldots	&	LT	\\   
2013 04 20	&	56402.2	&	\phs90.3	&	\ldots	&	\ldots	&	20.79 (0.12)	&	\ldots	&	\ldots	&	NTT	\\   
2013 12 29	&	56655.3	&	328.4	&	\ldots	&	\ldots	&	\ldots	&	25.0$^\dag$	&	\ldots	&	VLT	\\    
 \hline
\end{tabular}
\tablefoot{
*Epoch in days relative to the date of maximum brightness (MJD 56306.21) in the rest frame of the SN.
$^\dag$Synthetic magnitude determined from spectrum (see Table \ref{table:spectra_log}).
}
\end{table*}

\begin{table*}
\setlength{\tabcolsep}{3pt}
\caption{Optical $griz$ photometry of LSQ13fn. Uncertainties are given in parentheses. The magnitudes have not been corrected for reddening.}
\label{table:photometry_griz} 
 \centering
 \begin{tabular}{lccccccc}
 \hline\hline 
Date	&	MJD	&	Epoch*	&	$g$	&	$r$	&	$i$	&	$z$	&	Telescope	\\
yyyy mm dd	&	&	&	&	&	&	&	\\
\hline
2013 01 13	&	56305.3	&	\phs$-0.8$	&	\ldots	&	\ldots	&	19.78 (0.05)	&	\ldots	&	NTT	\\
2013 01 20	&	56312.3	&	\phs\phs5.7	&	\ldots	&	\ldots	&	19.43 (0.08)	&	\ldots	&	NTT	\\
2013 01 22	&	56314.3	&	\phs\phs7.6	&	\ldots	&	\ldots	&	19.56 (0.07)	&	\ldots	&	NTT	\\
2013 01 25	&	56317.2	&	\phs10.3	&	19.78 (0.14)	&	19.48 (0.08)	&	19.46 (0.10)	&	19.53 (0.22)	&	LT	\\
2013 01 26	&	56318.1	&	\phs11.2	&	19.97 (0.27)	&	19.33 (0.13)	&	19.44 (0.13)	&	19.44 (0.34)	&	LT	\\
2013 01 30	&	56322.1	&	\phs15.0	&	19.87 (0.36)	&	19.43 (0.27)	&	19.47 (0.17)	&	\ldots	&	LT	\\
2013 01 30	&	56322.3	&	\phs15.1	&	\ldots	&	\ldots	&	19.60 (0.07)	&	\ldots	&	NTT	\\
2013 02 08	&	56331.1	&	\phs23.4	&	20.08 (0.13)	&	19.60 (0.08)	&	19.56 (0.09)	&	\ldots	&	LT	\\
2013 02 21	&	56344.1	&	\phs35.7	&	20.30 (0.29)	&	19.59 (0.12)	&	19.41 (0.26)	&	\ldots	&	LT	\\
2013 02 23	&	56346.1	&	\phs37.5	&	\ldots	&	19.75 (0.15)	&	19.53 (0.14)	&	\ldots	&	LT	\\
2013 03 11	&	56362.0	&	\phs52.5	&	20.43 (0.16)	&	19.87 (0.09)	&	19.65 (0.08)	&	19.49 (0.23)	&	LT	\\
2013 03 18	&	56369.0	&	\phs59.0	&	20.52 (0.13)	&	19.60 (0.09)	&	\ldots	&	\ldots	&	LT	\\
2013 03 21	&	56372.1	&	\phs62.0	&	20.53 (0.17)	&	19.64 (0.30)	&	\ldots	&	\ldots	&	LT	\\
2013 03 25	&	56376.0	&	\phs65.6	&	20.57 (0.34)	&	19.70 (0.12)	&	19.60 (0.06)	&	19.74 (0.22)	&	LT	\\
2013 03 29	&	56380.1	&	\phs69.5	&	\ldots	&	19.95 (0.19)	&	19.79 (0.18)	&	\ldots	&	LT	\\
2013 03 31	&	56382.0	&	\phs71.3	&	20.68 (0.28)	&	19.82 (0.14)	&	19.60 (0.26)	&	\ldots	&	LT	\\
2013 04 14	&	56396.1	&	\phs84.5	&	\ldots	&	\ldots	&	19.75 (0.08)	&	\ldots	&	NTT	\\   
2013 04 15	&	56397.0	&	\phs85.4	&	21.23 (0.18)	&	20.04 (0.07)	&	19.62 (0.10)	&	19.80 (0.25)	&	LT	\\
2013 04 28	&	56409.9	&	\phs97.6	&	\ldots	&	20.25 (0.08)	&	\ldots	&	\ldots	&	LT	\\
2013 05 04	&	56415.9	&	103.2	&	\ldots	&	20.32 (0.06)	&	19.85 (0.10)	&	\ldots	&	LT	\\
2013 05 23	&	56434.9	&	121.1	&	\ldots	&	$>$21.35	&	$>$22.00	&	\ldots	&	LT	\\
2013 05 24	&	56435.9	&	122.0	&	\ldots	&	$>$22.18	&	\ldots	&	\ldots	&	LT	\\ 
\hline
\end{tabular}
\tablefoot{*Epoch relative to the date of maximum brightness (MJD 56306.21) in the rest frame of the SN.}
\end{table*}

\begin{table*}
\caption{Local $UBVRI$ sequence stars use to calibrate LSQ13fn photometry. Uncertainties are given in parentheses.}
\label{table:sequence_ubvri} 
 \centering
 \begin{tabular}{lccccccc}
 \hline\hline 
\# & RA & Dec & $U$ & $B$ & $V$ & $R$ & $I$ \\
 \hline
1	&	11:51:17.775	&	$-$29:34:54.09	&	18.93 (0.04)	&	19.14 (0.01)	&	18.11 (0.01)	&	17.20 (0.01)	&	16.10 (0.02)		\\
2	&	11:51:21.373	&	$-$29:35:17.58	&	15.31 (0.06)	&	15.24 (0.01)	&	14.58 (0.01)	&	14.18 (0.01)	&	13.81 (0.02)		\\
3	&	11:51:22.206	&	$-$29:35:25.91	&	18.60 (0.06)	&	18.46 (0.02)	&	17.69 (0.01)	&	17.20 (0.00)	&	16.72 (0.01)		\\
4	&	11:51:10.039	&	$-$29:35:06.90	&	19.18 (0.06)	&	19.25 (0.02)	&	18.57 (0.02)	&	18.13 (0.01)	&	17.66 (0.01)		\\
5	&	11:51:21.139	&	$-$29:36:38.81	&	20.53 (0.06)	&	19.47 (0.01)	&	18.01 (0.01)	&	17.06 (0.00)	&	16.07 (0.01)		\\
6	&	11:51:25.397	&	$-$29:37:43.70	&	19.95 (0.05)	&	18.90 (0.00)	&	17.45 (0.02)	&	16.51 (0.02)	&	15.47 (0.01)		\\
7	&	11:51:24.406	&	$-$29:37:59.94	&	18.39 (0.05)	&	18.48 (0.01)	&	18.08 (0.01)	&	17.80 (0.01)	&	17.44 (0.02)		\\
8	&	11:51:21.321	&	$-$29:37:53.01	&	18.96 (0.08)	&	17.72 (0.02)	&	16.52 (0.02)	&	15.76 (0.01)	&	15.06 (0.02)		\\
9	&	11:51:21.954	&	$-$29:38:12.99	&	17.73 (0.06)	&	17.36 (0.01)	&	16.62 (0.01)	&	16.21 (0.01)	&	15.77 (0.01)		\\
 \hline
\end{tabular}
\end{table*}

\begin{table*}
\caption{Local $griz$ sequence stars use to calibrate LSQ13fn photometry. Uncertainties are given in parentheses.}
\label{table:sequence_griz} 
 \centering
 \begin{tabular}{lcccccc}
 \hline\hline 
 \# & RA & Dec & $g$ & $r$ & $i$ & $z$ \\
 \hline
1	&	11:51:17.775	&	$-$29:34:54.09	&	18.64 (0.04)	&	17.26 (0.04)	&	16.29 (0.03)	&	15.61 (0.04)	\\
2	&	11:51:21.373	&	$-$29:35:17.58	&	14.86 (0.03)	&	14.16 (0.02)	&	13.91 (0.02)	&	13.78 (0.02)	\\
3	&	11:51:22.206	&	$-$29:35:25.91	&	18.08 (0.02)	&	17.17 (0.03)	&	16.83 (0.02)	&	16.60 (0.03)	\\
4	&	11:51:10.039	&	$-$29:35:06.90	&	18.83 (0.04)	&	18.05 (0.03)	&	17.78 (0.04)	&	17.68 (0.03)	\\
5	&	11:51:21.139	&	$-$29:36:38.81	&	18.74 (0.04)	&	17.09 (0.03)	&	16.27 (0.02)	&	15.65 (0.03)	\\
6	&	11:51:25.397	&	$-$29:37:43.70	&	18.16 (0.04)	&	16.55 (0.04)	&	15.68 (0.03)	&	15.06 (0.03)	\\
7	&	11:51:24.406	&	$-$29:37:59.94	&	18.28 (0.03)	&	17.71 (0.04)	&	17.55 (0.04)	&	17.39 (0.05)	\\
8	&	11:51:21.321	&	$-$29:37:53.01	&	17.13 (0.04)	&	15.75 (0.03)	&	15.20 (0.03)	&	14.79 (0.03)	\\
9	&	11:51:21.954	&	$-$29:38:12.99	&	16.96 (0.04)	&	16.16 (0.03)	&	15.85 (0.03)	&	15.67 (0.01)	\\
 \hline
\end{tabular}
\end{table*}

\begin{table*}
\setlength{\tabcolsep}{3pt}
\caption{Log of the spectroscopic observations of LSQ13fn} 
\label{table:spectra_log} 
\centering 
\begin{tabular}{lclccccc} 
\hline\hline 
Date	&	MJD	&	Epoch*	&	Telescope + intstrument	&	Wavelength range	&	Exposure time	&	Slit width	&	Resolution$^{\dag}$	\\
yyyy mm dd	&		&		&		&	(\AA)	&	(s)	&	(")	&	(\AA)	\\
\hline															
2013 01 13	&	56305.3	&	\phs$-$0.9	&	NTT+EFOSC2+gr13	&	3650$-$9250	&	1$\times$2400	&	1.0	&	17.7	\\
2013 01 20	&	56312.3	&	\phs\phs5.7	&	NTT+EFOSC2+gr11	&	3350$-$7460	&	1$\times$3600	&	1.0	&	14.3	\\
2013 01 22	&	56314.2	&	\phs\phs7.6	    &	NTT+EFOSC2+gr11	&	3350$-$7460	&	2$\times$2400 + 1$\times$1800	&	1.0	&	13.9	\\
2013 01 30	&	56322.3	&	\phs15.1	&	NTT+EFOSC2+gr11	&	3380$-$7520	&	2$\times$2400	&	1.0	&	14.4	\\
2013 02 09	&	56332.2	&	\phs24.4	&	NTT+EFOSC2+gr11	&	3350$-$7460	&	3$\times$2220	&	1.0	&	14.4	\\
2013 03 03	&	56354.1	&	\phs45.0	&	NTT+EFOSC2+gr13,16	&	3650$-$10000	&	2$\times$1800, 2$\times$1800	&	1.0	&	18.6, 13.0	\\
2013 03 13	&	56364.3	&	\phs54.6	&	NTT+EFOSC2+gr11,16	&	3350$-$10000	&	2$\times$2400, 2$\times$2400	&	1.0	&	14.1, 13.8	\\
2013 03 17	&	56368.2	&	\phs58.3	&	NTT+EFOSC2+gr11,16	&	3350$-$10000	&	2$\times$2400, 2$\times$2400	&	1.0	&	14.0, 13.2	\\
2013 03 30	&	56382.0	&	\phs71.3	&	WHT+ISIS+R300B,R158R	&	3250$-$10000	&	2$\times$1800, 2$\times$1800	&	1.0	&	4.5, 6.0	\\
2013 04 05	&	56387.2	&	\phs76.2	&	NTT+EFOSC2+gr13 	&	3650$-$9250	&	3$\times$2700	&	1.0	&	19.0	\\
2013 04 20	&	56402.2	&	\phs90.3	&	NTT+EFOSC2+gr13 	&	3650$-$9250	&	3$\times$2400	&	1.0	&	19.0	\\
2013 12 29	&	56655.3	&	328.4	&	VLT+FORS2+300V	&	4350$-$9600	&	2$\times$1100	&	1.0	&	\phs8.2	\\
\hline 
\end{tabular}
\tablefoot{
*Epoch relative to the date of maximum brightness (MJD 56306.21) in the rest frame of the SN. 
$^{\dag}$Resolution, as measured from the FWHM of the sky lines.
Note that the NTT spectra at 5.7\,d and 15.1\,d are not presented in Fig. \ref{fig_spectra_sequence} as they were taken in poor conditions and the supernova signal is weak.
}
\end{table*}

\begin{table*}[!t]
\setlength{\tabcolsep}{3pt}
\caption{Properties and references of SNe referred to in this paper.}
\label{table:SN_properties} 
 \centering
 \begin{tabular}{lccccc}
 \hline\hline 
SN	&	SN type & v$_{hel}$	&	$\mu$	&	$E(B-V)$	&	Reference	\\
	&	& (\kms)	&	(mag)	&	(mag)	&		\\
\hline									
LSQ13fn	&	II & 18900	&	37.10	&	0.054	&		\\
1998S	& IIL/IIn	& \phs\phs895*	&	31.18	&	0.230	&	1, 2, 3	\\
1999br	& IIP &	\phs\phs960*	&	31.19	&	0.024	&	4	\\
1999em	& IIP &	\phs\phs717*	&	30.34	&	0.100	&	5, 6, 7	\\
2004et	&	IIP & \phs\phs\phs40*	&	28.85	&	0.410	&	8, 9	\\
2005cs	& IIP &	\phs\phs600*	&	29.26	&	0.050	&	10	\\
2007pk	& IIL/IIP/IIn &	\phs4993*	&	34.23	&	0.110	&	11, 12	\\
2009bw	&	IIP	&	\phs1155*	&	31.53	&	0.310	&	13 \\
2010al$^{\dag}$	&	Ibn & \phs5143*	&	34.27	&	0.057	&	14	\\
2013cu$^{\dag}$	&	IIb & \phs7565*	&	35.17	&	0.011	&	17	\\
 \hline
\end{tabular}
\tablefoot{
*Taken from NASA/IPAC Extragalactic Database (NED). 
$^{\dag}$Spectra downloaded from Weizmann Interactive Supernova Data Repository (WISeREP; \citealt{Yaron_2012}; \url{http://www.weizmann.ac.il/astrophysics/wiserep/}). 
1) \citet{liu_2000}; 
2) \citet{Baron_2000}; 
3) \citet{fassia_2001}; 
4) \citet{Pastorello_2004}; 
5) \citet{hamuy_2001}; 
6) \citet{elmhamdi_2003};
7) \citet{leonard_2003};
8) \citet{sahu_2006}; 
9) \citet{maguire_2010}; 
10) \citet{Pastorello_2009}; 
11) \citet{Pritchard_2012};
12) \citet{Inserra_2013}; 
13) \citet{Inserra_2012}; 
14) \citet{Pastorello_2015}; 
16) \citet{Gal-Yam_2014}.
}
\end{table*}

\end{document}